\documentclass[10pt,a4paper]{article}

\usepackage[margin=1.3in]{geometry}
\usepackage[utf8]{inputenc}
\usepackage[english]{babel}
\usepackage{amsmath}
\usepackage{amsfonts}
\usepackage{amssymb}
\usepackage{bbold}
\usepackage[title]{appendix}
\usepackage{graphicx}
\author{Valerio Astuti, Silvio Franz, Giorgio Parisi}
\title{New analysis of the free energy cost of interfaces in spin glasses}
\date{}
\begin{document}
\maketitle
\begin{abstract}
In this work we want to enhance the calculation performed by Franz, Parisi and Virasoro (FPV) \cite{Franz92, Franz94} to estimate the free energy cost of interfaces in spin glasses and evaluate the lower critical dimension at which replica symmetry is restored. In particular we evaluate the free energy cost for a general class of effective Hamiltonians showing full replica symmetry breaking, and study the dependence of this cost on the order parameter and on the temperature. We confirm the findings of the FPV papers for the scaling of the free energy, recovering a value for the lower critical dimension of $D_{lc} = 2.5$. In addition to their results we find a non-trivial dependence of the free energy density cost on the order parameter and the temperature. Apart from the case of a restricted class of effective Hamiltonians this dependence cannot be expressed in terms of functions with a clear physical interpretation, as is the case in hierarchical models \cite{Franz2009}. In addition we connect the results on the lower critical dimension with recent simulations \cite{Maiorano18}.
\end{abstract}

\section{Introduction}
A very powerful tool to study spontaneous symmetry breaking is the evaluation of the free energy cost of interfaces between different ordered phases of a system. The stability of the ordered phase is in fact strictly linked to the probabilistic cost of fluctuations transforming one ordered phase into another. If the probability of creating bubbles of a different ordered phase inside a given one is not zero the ordered phase will be unstable under this kind of fluctuations, so in the end no ordered phase can exist \cite{Zinn96}. 
The free energy cost of interfaces between different phases will usually have a strong dependence on the number of space dimensions, so if we are able to evaluate this cost for a generic number of dimensions we can in turn determine if a low-temperature ordered phase is to be expected for a given dimensionality of space.

This technique can be used to easily determine the so called \emph{lower critical dimension} of systems composed of Ising and continuous (Heisenberg) spins. 
As the name suggests the lower critical dimension is the number of spatial dimensions for which the free energy cost of fluctuations cease to be infinite in the thermodynamic limit, such that in this number of dimension spontaneous symmetry breaking is not possible. 

The formalization of these concepts is the \emph{Mermin-Wagner-Hohenberg theorem} \cite{Zinn96}, which links the existence of spontaneous symmetry breaking with the number of dimensions and the symmetries of the system under study. The theorem states that if a continuous symmetry is present in the Hamiltonian of the system then no ordered phase can exist in dimension $D=2$ or lower. The continuous symmetry implies the absence of an energy gap in the fluctuations around the ordered phase, and the presence of Goldstone modes in the spontaneously broken phase.
These modes will be strong enough to destroy the ordered phase as long as the number of spatial dimension is equal or lower than two, as can be easily seen even by dimensional analysis. By virtue of this theorem the Heisenberg spin system, for which a rotational symmetry is valid, must have lower critical dimension $D_{lc} = 2$.

This result can be easily sketched by considering a $D-$dimensional cube of side $L$ with periodic boundary conditions in the first $D-1$ directions, and boundary conditions in the $D$-th direction given by spins displaced by an angle $\theta$. The difference in free energy at low temperature between this setting and the one with only periodic boundary conditions will be given by the interaction energy between neighbouring spins, which in turn are forced to be different by the different boundary conditions. The least energy will be used when every couple of neighbouring spins along the $D$-th direction is displaced by an angle $\frac{\theta}{L}$. This displacement implies an energy density proportional to $\frac{\theta^2}{L^2}$, and so the free energy difference with respect to the periodic boundary conditions system is proportional to $\theta^2 L^{D-2}$.
We have to deal with a different situation in systems with only discrete symmetries. In such cases we indeed have an energy gap between two different ordered phases, so no cheap fluctuations will be present. In the situation presented above the least energy configurations are not the ones in which the spin difference is distributed over the whole system, but the ones in which the interface is concentrated on a single hypersurface (in this case there is a lower bound in the interaction energy, such that there is no point in distributing it over larger regions of the system). This implies a free energy cost proportional to $L^{D-1}$, and a lower critical dimension $D_{lc}=1$.  

The system of interest in this paper is the Ising spin glass, or Edwards-Anderson model, and as we will see it will escape both the situations described above. 
The Hamiltonian of this system contains a quenched disorder - unknown variables which however do not change from one configuration to another - in addition to Ising spin variables. For a fixed quenched disorder the only symmetry of this Hamiltonian is the one of the Ising spins, but in the end we will have to average out the disorder variables, introducing the so-called \emph{replicas} of the system \cite{mezard1987spin}. They are copies of the system in which the quenched disorder is identical but the spins configurations are different, and by construction they ought to be equivalent once we average over the spins configuration. The order parameter of the replicated system is the \emph{overlap} between replicas, and this treatment implies a new symmetry for the effective Hamiltonian of this system: the symmmetry under exchange of replicas. This symmetry can be described as continuous, and the zero modes associated with it are even more pervasive than the ones described by the Mermin-Wagner-Hohenberg theorem. These modes are powerful enough to destroy the ordered phase at dimension higher than two, and the value of the lower critical dimension for this system - though still debated - is greater than the one for Heisenberg spins. 

While the presence of a stable phase in three spatial dimensions is strongly supported by experiments and numerical simulations \cite{Maiorano18, PhysRevB.43.8199, PhysRevLett.82.5128, PhysRevB.62.14237} there is still no definitive theoretical result \cite{Fisher86, bray1987scaling}. 
In two related works \cite{Franz92, Franz94} Franz, Parisi and Virasoro managed to derive a value for the lower critical dimension based on the replica mean field formulation of the problem. Their calculation is done for the so-called truncated model \cite{mezard1987spin, bray1978replica, pytte1979scaling, parisi1980order}, an expansion of the mean field effective Hamiltonian for temperatures close to the critical one. They impose a difference $\Delta q$ in the overlap order parameter over a distance $L$ in one spatial direction, keeping free boundary conditions in the other directions.
The main result of their papers is a free energy cost which scales as $\delta F \propto L^{D - \frac{5}{2}}$ where $L$ is the linear dimension of the system. This implies a lower critical dimension $D_{lc} = \frac{5}{2}$, as for this dimension and below we obtain a finite probability for the coexistence of different phases. In addition the free energy cost is proportional to $|\Delta q|^{\frac{5}{2}}$, which implies typical fluctuations of length $\ell$ with magnitude of order $|\Delta q\left( \ell	\right)|_{typ}\sim \ell^{1 - \frac{2 D}{5}}$. 
These results, though confirmed even by recent simulations \cite{Maiorano18}, suffers from two major shortcomings: the first is that it is based on mean field theory; the second is the limitation to the truncated model. 
Mean field theory is a correct description of the Edwards-Anderson model when the number of spatial dimension $D$ is equal or greater than 6, but its validity is debated in lower dimensions. It is obviously exact in the fully connected version of the Edwards-Anderson model, also called Sherrington-Kirkpatrick model. The exact solution for this model was found at the beginning of the eighties by one of the authors, and it first unveiled the existence of the spin glass phase, in which the symmetry under exchange of replicas is broken in a hierarchical pattern \cite{mezard1987spin, parisi1980order}. 
In ordered systems we can reliably extend our knowledge outside the domain of mean field theory by means of the renormalization group. In spin glass systems with full replica symmetry breaking however a renormalization group treatment is much more cumbersome due to the complex nature of the order parameter. In addition in an $\epsilon$ expansion we find strongly divergent series, which render any result unreliable even in dimension $D=5$. In this situation we don't have strong evidence that mean field theory results are valid down to the lower critical dimension, and we have to rely heavily on simulation and experiments for confirmation of our results. 
The use of the truncated model imposes less severe limitations than the mean field theory approximation. In the evaluation of the free energy cost it implies we are retaining only the lowest order approximation in the difference between the temperature $T$ and the critical temperature $T_c$, also called the \emph{reduced temperature} and denoted by $\tau$. In particular we cannot have any dependence on the order parameter at this order of approximation, and in the FPV paper the free energy cost is constant as a function of the overlap.
An interface calculation similar to the one in FPV was performed in \cite{Franz2009} for hierarchical spin-glass models, a slightly simpler system.
In this models the spatial structure of the Edwards-Anderson model is replaced with a binary tree of which the leaves are occupied by spins. The distance between two such spins is $2^k$, where $k$ is the number of branching from a common node. The Hamiltonian is costructed iteratively following the law (here the $J_{ij}$ follow as usual the normal distribution with zero average and unit variance and $\sigma$ is a parameter tuning the strength of the interaction):
\begin{equation}
H_{k+1}^J \left[S_1, ..., S_{2^{k+1}}  \right] = H_{k}^{J_1} \left[S_1, ..., S_{2^{k}}  \right] + H_{k}^{J_2} \left[S_{2^k+1}, ..., S_{2^{k+1}}  \right] - \frac{1}{2^{(k+1)\sigma}} \sum_{i<j}^{2^{k+1}} J_{ij} S_i S_j
\end{equation}
and the starting condition:
\begin{equation}
H^J_1 \left( S_1, S_2\right) = - \frac{J}{2^{\sigma}} S_1 S_2
\end{equation}
For this system it was possible to study the dependence of the free energy cost on the order parameter. The simple proportionality $\delta f\left(q \right) \propto P\left(q \right)$ was found, where $\delta f(q)$ is the free energy density cost as a function of the order parameter $q$ and $P\left(q\right)$ is its probability distribution. 
More precisely one can evaluate the probability of a fluctuation bringing a state having the same boundary overlap $p_1$ to a different state, having different boundary overlaps $p_1$ and $p_2$. The probability one obtains has the form:
\begin{equation}
\rho_{k+1}\left(p_1, p_2 \right) \propto e^{-P\left(\frac{p_1 + p_2}{2}\right)2^{2(1-\sigma)(k+1)}|p_1-p_2|^3}
\end{equation}
One might wonder if this kind of dependence from $p_1$ and $p_2$ is general or just a peculiarity of hierarchical spin glasses, but to study the problem in the Edwards-Anderson model we have to go past the approximations used in the FPV papers. With these approximations the quantity $P\left(q \right)$ is a constant, so that the free energy cost is trivially proportional to it, but when the probability is non-trivial the proportionality could be broken.   

In addition a recent work \cite{Maiorano18} simulated the cost in energy of interfaces in the Edwards-Anderson model as a function of the number of space dimensions.
Also for the comparison with this result we need to go past the approximation of the truncated model, in order to be able to describe the temperature dependence of the free energy.

The main purpouse of this paper is to address the last two issues, extending the FPV calculation in order to describe the dependence of the free energy cost on the temperature and the order parameter, in addition to the number of space dimensions.
The paper is organized in this way: in section \ref{trunc model section} we give the basic definition of the model, mostly borrowing from the original paper \cite{Franz94}. 
In section \ref{varappr} we present the variational approach to evaluate the free energy cost, and in section \ref{highorder} we show how the previously obtained result is essentially stable in the space of possible variations of the solution.
In the remaining sections of the paper we generalize the model to evaluate the effect of non-linearities in the solution of the equations of motion on the free energy cost of the interface.

\section{The FPV computation}
\label{trunc model section}
The starting model of the FPV paper was the $D$-dimensional Edwards-Anderson spin-glass Hamiltonian in a box of volume $V$ and side $L \gg 1$:
\begin{equation}
H\left(\{s\}\right)=- \sum_{<i,j>} J_{ij} s_i s_j - h \sum_i s_i
\end{equation} 
where $<i,j>$ are nearest neighbour sites and the spins $s_i$ are of the Ising type, $s_i = \pm 1$. The variables $J_{ij}$ are independent Gaussian variables such that\footnote{We indicate with $\overline{\cdot}$ the average over the disorder and with $\langle \cdot \rangle$ termodynamic averages.} $\overline{J_{ij}} = 0$ and $\overline{J^2_{ij}} = J^2 = 1$. 
The model is studied close to the critical temperature $T_c$, with a mean field approach.

%To evaluate the free energy cost of an interface the method of real replicas {\bf[cite FPV 92]} was used. We consider two copies of the same system (same quenched disorder) described by the total hamiltonian:
%\begin{equation}
%H_2\left(\{s^a, s^b \}\right) = H\left(\{s^a\} \right) + H\left(\{s^b \}\right)
%\end{equation}
%
%The order parameter in the replica framework is the position dependend overlap matrix $Q_i^{ab} = \langle s_i^a s_i^b	\rangle$. The free energy as a function of the order parameter can be expressed as a Landau expansion coarse graining the lattice, and the space dependent matrix field becomes $Q_{ab}(x) = \frac{1}{|V_x|}\sum_{i \in V_x}Q_i^{ab}$. 
%Given we are close to the critical temperature we can truncate the Landau expansion to the fourth power of the field and keep only the fourth order term responsible for the replica symmetry breaking, defining the so called \emph{truncated model}. The coarse grained free energy has the form:

Taking the continuum limit and considering the system close to the critical temperature we obtain the expression for the mean field free energy:
\begin{equation}
\label{truncated}
-2 n F = \int d^D x \Bigg[ \texttt{Tr}(|\nabla Q (x)|^2)  + \tau \texttt{Tr} Q^2(x) + \frac{1}{3} 
\texttt{Tr} Q^3(x) + \frac{y}{4}\sum_{ab}Q_{ab}(x)^4 + h^2 \sum_{ab} Q_{ab}(x) \Bigg]
\end{equation}
%\begin{multline}
%\label{truncated}
%-2 n F = \int d^D x \Bigg[ \texttt{Tr}(|\nabla Q (x)|^2)  + \tau \texttt{Tr} Q^2(x) + \frac{1}{3} 
%\texttt{Tr} Q^3(x) +  \\ + \frac{y}{4}\sum_{ab}Q_{ab}(x)^4 + h^2 \sum_{ab} Q_{ab}(x) \Bigg]
%\end{multline}
Here $Q_{ab}(x)$ is the overlap matrix, the order parameter of the replica theory. Considering two replicas of the system - having the same disorder but in general different configurations - we can define the overlap matrix as $Q_{ab}^i = \langle s_a^i s_b^i \rangle$, where $a$ and $b$ are the replica indices, and $i$ is the site index. In the continuum limit , fixing a small region of space $V_x$ centered in $x$ and having volume $|V_x|$, this becomes $Q_{ab}(x) = \frac{1}{|V_x|}\sum_{i \in V_x}Q_{ab}^i$. 
In all the paper $\tau = T_c - T$, $y$ is the coupling constant of the replica symmetry breaking interaction (it is equal to $\frac{2}{3}$ in the original model), $\texttt{Tr}$ is the trace in replica space, and $n$ is the number of replicas. In the FPV computation the only quartic term retained was the one responsible for the replica symmetry breaking. The Edwards-Anderson model to which the above approximations are applied is also called the \emph{reduced model}.

The reduced model with free boundary conditions can be solved by the mean field solution found by Parisi \cite{mezard1987spin}. In this solution the symmetry between replicas - present in the effective Hamiltonian by construction - is broken in an infinite, hierarchical way. To describe this solution a limit for the number of replicas $n \to 0$ is needed, and the pair of replica indices $ab$ is replaced by a continuous codistance in replica space. At the saddle point the solution is an order parameter constant in space, with a replica space dependence of the form:
\begin{equation}
Q_{ab}(x) \to q\left(x , u\right) = q\left( u\right) = \begin{cases} q_{\text{min}}\qquad u \leq u_0  \\  \frac{u}{3 y}\qquad u_0 \leq u \leq u_1  \\ q_{\text{max}}\qquad u \geq u_1 \end{cases}
\end{equation}
with $u_0 = 3 y q_{\text{min}}$ and $u_1 =  3 y q_{\text{max}}$.
Here and in the rest of the paper we use the letter $x$ for \textbf{space} coordinates, and $u$ for the \textbf{replica} coordinates. 

For the breaking points we have the relations:
\begin{equation}
2 y q_{\text{min}}^3 = h^2 
\end{equation}
\begin{equation}
q_{\text{max}}\left( 1 - \frac{3 y }{2} q_{\text{max}} \right) = \tau
\end{equation}
The solution found predicts many overlapping pure states, with an overlap constant in space. 
From this starting point we want to force a dishomogeneity in the order parameter by imposing different boundary conditions over a distance $L$ in a particular direction of the lattice. 
We cannot directly impose particular values for $q(0,u)$ and $q(L,u)$ because the functional form $q(x, u)$ is obtained from the saddle point equations, so any modification of it would bring us out of the saddle point approximation. 
Instead the method of real replicas is used \cite{Franz92, Franz94}. 
Two copies of the system are considered, with the same disordered couplings $J_{ij}$, the same temperature and magnetic field, but different thermal configurations $\{ s^i_a \}$ and $\{ s^i_b \}$. Being the two boundaries on which we force different conditions $B_1$ (on which $x=0$) and $B_2$ (on which $x=L$), the partition function of the constrained system can be written as:
\begin{equation}
Z_{p_1,p_2} = \sum_{\{s^i_a,s^i_b\}}e^{-\beta H_J\left[s^i_a\right]- \beta H_J\left[s^i_b\right]} \prod_{x\in B_1}\delta\left(Q_{ab}(x) - p_1 \right)  \prod_{x\in B_2} \delta\left(Q_{ab}(x) - p_2 \right)
\end{equation}
As it can be seen from the above definition the overlap boundary conditions are constrained only in one spatial direction, the other ones remaining free. The overlap will thus be a constant in all directions but the constrained one.
The quantity we are interested in is the free energy difference corresponding to the ratio between this partition function and the one without constraints. 
We have to remember that the replicas used to obtain the mean field solution of the problem are conceptually different from the replicas we are using to evaluate the free energy cost of the constraint. For this reason the latter are called \emph{real replicas}, and when applying the usual replica method to the coupled system we have two different replica indices, one for each real replica.    
In this setting a generalized order parameter is used, in that we have to consider the overlaps between both two copies of the first (or the second) real replica, and the overlap between the first and the second real replicas.
Thus we have three different types of overlaps: $Q^{ab}_{11}(x)$ between states of the first real replica, $Q^{ab}_{22}(x)$ between states of the second real replica, and $Q^{ab}_{12}(x)=Q^{ba}_{21}(x)$ between states of the first and states of the second real replica.  
In the original work the ansatz taken is $Q^{ab}_{11}(x)=Q^{ab}_{22}(x)=Q_{ab}(x)$, $Q^{ab}_{12}(x)=Q^{ab}_{21}(x)=P_{ab}(x)$ and $Q^{aa}_{12}(x)=\tilde{p}(x)$ ($Q_{aa}(x)=0$ is taken by convention). Here $\tilde{p}(x)$ is an unknown function we have to recover from the saddle point equations with boundary conditions $\tilde{p}(0)=p_1$ and $\tilde{p}(L)=p_2$. 

This generalized order parameter can be written as $Q_{\alpha \beta}(x)$, $\alpha$ and $\beta$ being double indices: $\alpha = \{a,r\}$, $\beta = \{b,s\}$ for $r$, $s=1$, $2$. 
We will consider different boundary conditions only in one spatial direction, so here the coordinate $x$ will denote the only direction in which a non-trivial behaviour of the system is present. 

Near the critical temperature a Landau expansion of the free energy as a function of this extended order parameter can be performed, and in the case $p_1 = p_2$ a saddle point solution can be found with similar methods as in the single real replica problem.  
The solution has the form:
\begin{equation}
Q_{ab}(x) \to q(x,u) = q(u) = \begin{cases} q_{\text{min}} \qquad & 0\leq u \leq \frac{u_0}{2}  \\ \frac{2 u }{3 y} \qquad & \frac{u_0}{2} <  u  \leq \frac{u_p}{2} \\ \tilde{p} \qquad & \frac{u_p}{2} <  u \leq u_p  \\ \frac{u}{3 y} \qquad  & u_p <  u \leq u_1  \\ q_{\text{max}} \qquad & 	u_1 <  u \leq 1 \end{cases}
\end{equation}
\begin{equation}
P_{ab}(x) \to p(x,u) = p(u) = \begin{cases} q_{\text{min}} \qquad & 0\leq u \leq \frac{u_0}{2} \\ \frac{2 u }{3 y} \qquad & \frac{u_0}{2} <  u  \leq \frac{u_p}{2} \\ \tilde{p} \qquad & \frac{u_p}{2} <  u \leq 1  \end{cases}
\end{equation}
\begin{equation}
\tilde{p}(x) = \tilde{p} = p_1 = p_2
\end{equation}
The parameters are the same as in the unconstrained solution, except for $u_p = 3 y \tilde{p}$.  
The solution of the unconstrained problem being $Q_f(u) = \frac{u}{3 y}$, we see that in the region $u<u_p$ the solution can be written as $Q_f(2 u)$. This is a general fact, not based on the particular model but on ultrametric symmetry.
We can now perturb this solution of the problem with $p_1 = p_2$ to study the case $p_1 \neq p_2$. The free energy density is modified by the gradient squared $|\nabla Q|^2$, and the parameter of the perturbation will be proportional to some positive power of $|p_1 - p_2|/L$ (the perturbation in density must go to zero for fixed overlap difference and infinitely far away boundaries).
The gradient term is the one responsible for the value of the lower critical dimension $D_{lc} = 2$ in the presence of a continuous symmetry. It is easy to see that it gives a contribution to the free energy cost proportional to $L^{D-2}$, which would guarantee the stability of ordered phase down to $D=2$. In a spin glass system however this dominant contribution vanishes due to replica symmetry. In fact for the gradient term evaluated on the unperturbed saddle point we obtain:
\begin{multline}
\int d^D x \; \texttt{Tr}(|\nabla Q (x)|^2) = \\
= L^{D-1} \int dx \; \left( \frac{d \tilde{p}}{d x} \right)^2 \left[1 - \int du \; \theta\left(u  - \frac{u_p}{2} \right) \theta\left(u_p  - u \right)   - \int du \; \theta\left(u  - \frac{u_p}{2} \right)  \right] = 0
\end{multline} 
The vanishing of the gradient term at the leading order is a signal of the fact that breaking the replica symmetry is associated with zero modes which are much more powerful than the ones of a standard continuous symmetry breaking. 

Given that the gradient on the unperturbed solution vanishes we can conclude that the lower critical dimension will be higher than $D=2$. To estimate it an assumption on the functional form of the order parameter can be made, and an approximation of the saddle point can be found in the functional class selected. 
If we assume the function $\tilde{p}(x)$ to interpolate linearly between the two boundary condition, the saddle point equations evaluated on the unperturbed solution become:
\begin{eqnarray}
\label{saddlepoint}
\frac{\partial^2 q(x,u)}{\partial x^2} &=& 3 y \left(\frac{d \tilde{p}}{d x} \right)^2 \left[ \delta(u- u_p) - \frac{1}{2} \delta \left(u - \frac{u_p}{2}\right)\right] \\
\frac{\partial^2 p(x,u)}{\partial x^2} &=& - \frac{3 y}{2} \left(\frac{d \tilde{p}}{d x} \right)^2  \delta \left(u - \frac{u_p}{2}\right)
\end{eqnarray}
%\begin{equation}
%\label{saddlepoint1}
%\frac{\partial^2 q(x,u)}{\partial x^2} = 3 y \left(\frac{d \tilde{p}}{d x} \right)^2 \left[ \delta(u- u_p) - \frac{1}{2} \delta \left(u - \frac{u_p}{2}\right)\right]
%\end{equation}
%\begin{equation}
%\label{saddlepoint2}
%\frac{\partial^2 p(x,u)}{\partial x^2} = - \frac{3 y}{2} \left(\frac{d \tilde{p}}{d x} \right)^2  \delta \left(u - \frac{u_p}{2}\right)
%\end{equation}
Given the form of these equations the assumption is made that the effect of the gradient is to enforce a small smoothing of the overlap function around the points $\frac{u_p}{2}$ and $u_p$ in replica space. This assumption can be taken as the starting point of a variational problem, in which a polynomial function is chosen to smooth the unperturbed solution around the breaking points $\frac{u_p}{2}$ and $u_p$.

\begin{figure}[ht]
\centering
		\includegraphics[scale=0.8]{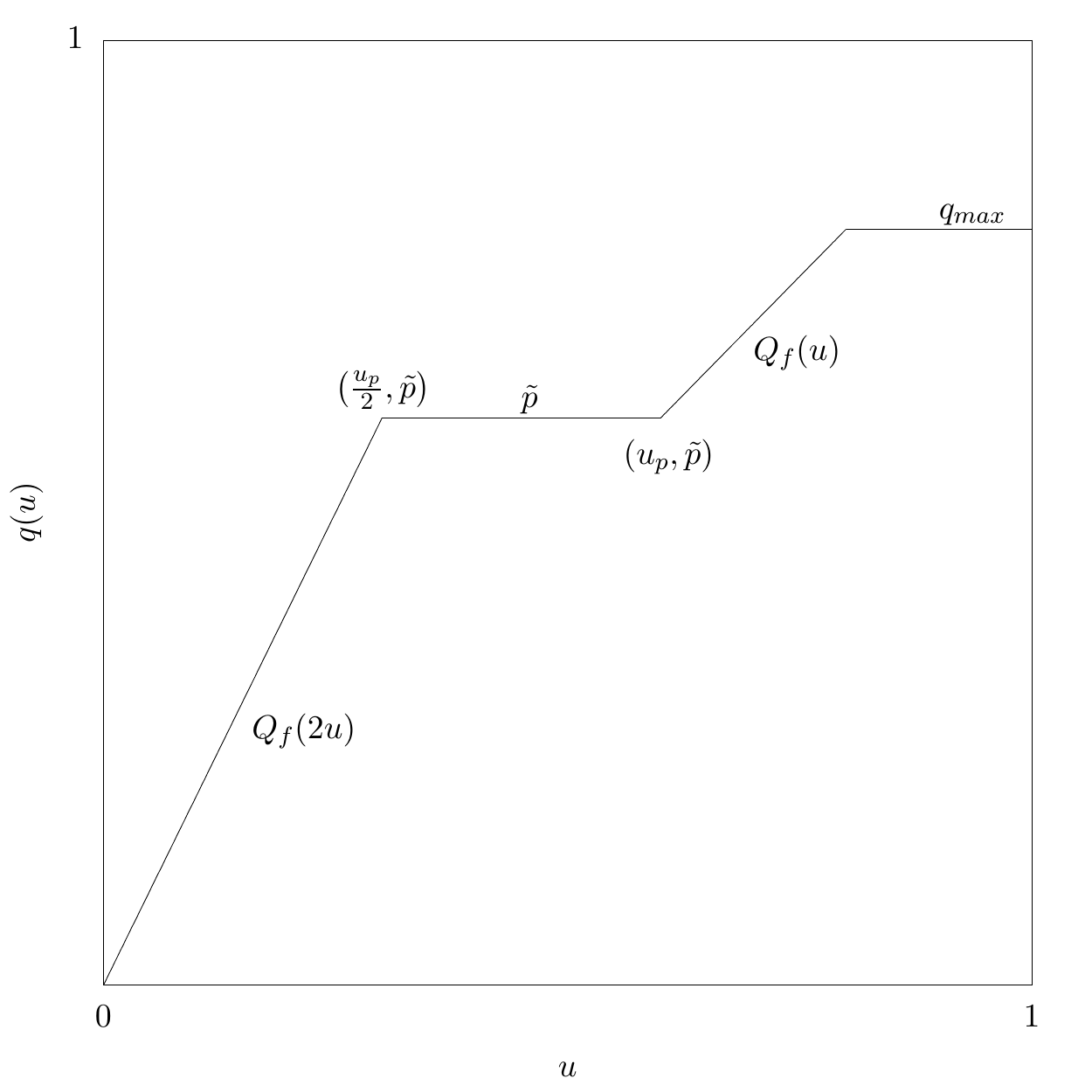}
\caption{Unperturbed same real-replicas overlap $q(u)$.}
\end{figure}

\begin{figure}[ht]		
\centering
		\includegraphics[scale=0.8]{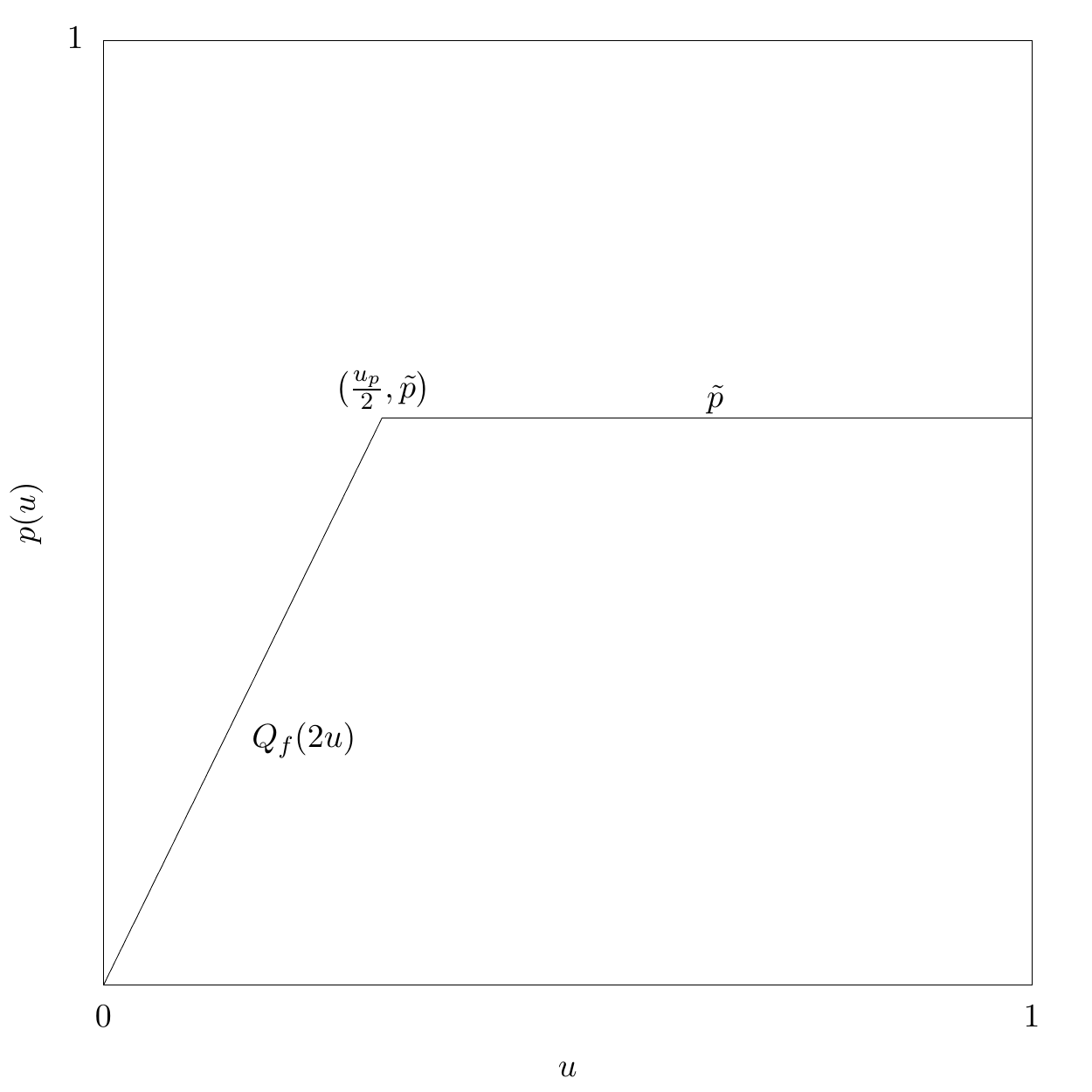}
\caption{Unperturbed different real-replicas overlap $p(u)$.}
\end{figure}

Before going to the variational problem in the next section, we mention a useful result about the dependence of the free energy density variation on the temperature and the magnetic field. In particular we can prove that as long as the unperturbed solutions $q(x,u)$ and $p(x,u)$ do not depend on the temperature and magnetic field in the region $u_0 < u < u_1$, the free energy variation $\delta F = F(p_1, p_2) - F(p_1, p_1)$ does not depend on these parameters either.
In fact we can expand the free energy variation as:
\begin{multline}
-2 n \delta F = \int d^D x \Bigg[ \texttt{Tr} \Big(|\nabla \delta Q (x)|^2\Big) + 2 \texttt{Tr} \Big(\nabla Q(x) \nabla \delta Q (x)\Big)  +  \tau \texttt{Tr} \Big( \delta Q^2(x)\Big) + \\ + \frac{1}{3} \texttt{Tr} \Big( \delta Q^3(x)\Big)  + \texttt{Tr} \Big( Q(x) \delta Q^2(x)\Big)  + \frac{y}{4}\sum_{ab}\Big(\Big( Q_{ab}(x) + \delta Q_{ab}(x) \Big)^4 -Q_{ab}^4(x) \Big)  \Bigg]
\end{multline}
with $\delta Q$ being a small perturbation localized around the points in which $q(u) = \tilde{p}$.
The only part of $\delta F$ which could depend on the temperature is 
\begin{equation}
\delta F_{\tau} = \tau \texttt{Tr} \Big( \delta Q^2(x)\Big) + \texttt{Tr} \Big( Q(x) \delta Q^2(x)\Big) 
\end{equation}
because all other terms in which $Q(x)$ is coupled with $\delta Q(x)$ are null when $Q(x)$ depend on $\tau$. It is however easy to show that also the temperature dependence of the two terms in $\delta F_{\tau}$ is such that the total dependence vanish. 
Similarly the only term in which we can find a dependence on the magnetic field is $\texttt{Tr} \Big( Q(x) \delta Q^2(x)\Big)$, but in all the points in which $Q(x)$ shows a dependence on the magnetic field $\delta Q^2(x)$ vanish, so also this term is field-independent.

\section{Variational approach}
\label{varappr}

In this section we show how a variational approach can be used to find an approximate solution to the saddle point equations \eqref{saddlepoint}. 
To avoid singularities in the equation of motion we want a solution which is derivable at the breaking points $\frac{u_p}{2}$, $u_p$. 
We smooth the unperturbed solution interpolating it with a quadratic polynomial on each of the breaking points, and maximize the free energy - and thus the free energy variation - varying the interval over which the unperturbed solution is smoothed.
%We smooth the unperturbed solution interpolating it with a polynomial (in the original paper a second order polynomial was used, here we extend the procedure to higher orders), and maximize the free energy - and thus the free energy variation - varying the interval over which the unperturbed solution is smoothed.
The family of solutions proposed in the FPV calculation has the form:
\begin{equation}
q(x,u) = \begin{cases} q_{\text{min}} \qquad & 0\leq u \leq u_0  \\ \frac{2 u }{3 y} \qquad & u_0 <  u  \leq u_1 \\  \tilde{p}(x)-\frac{(u - u_2)^2}{3 y \delta} \qquad & u_1 <  u \leq u_2  \\ \tilde{p}(x) \qquad & u_2 <  u \leq u_3  \\
\tilde{p}(x) + \frac{(u - u_3)^2}{6 y \delta'} & u_3 <  u \leq u_4 \\ \frac{u}{3 y} \qquad  & u_4 <  u \leq u_5  \\ q_{\text{max}} \qquad & 	u_5 <  u \leq 1 \end{cases}
\end{equation}
\begin{equation}
p(x,u) = \begin{cases} q_{\text{min}} \qquad & 0\leq u \leq u_0  \\ \frac{2 u }{3 y} \qquad & u_0 <  u  \leq u_1 \\  \tilde{p}(x)-\frac{(u - u_2)^2}{3 y \delta} \qquad & u_1 <  u \leq u_2  \\ \tilde{p}(x) \qquad & u_2 <  u \leq 1   \end{cases}
\end{equation}
\begin{equation}
p_1 \leq \tilde{p}(x) \leq p_2
\end{equation}
where the intervals are defined by the points:
\begin{equation}
\begin{aligned} 
u_0 = \frac{3 y q_{\text{min}}}{2} \qquad  u_1 = \frac{u_p - \delta}{2} \qquad  u_2 = \frac{u_p +  \delta}{2} \\  u_3 = u_p - \frac{\delta'}{2} \qquad   u_4 = u_p + \frac{\delta'}{2} \qquad  u_5 = 3 y q_{\text{max}}
\end{aligned}
\end{equation}
\begin{figure}[ht]
\centering
		\includegraphics[scale=0.8]{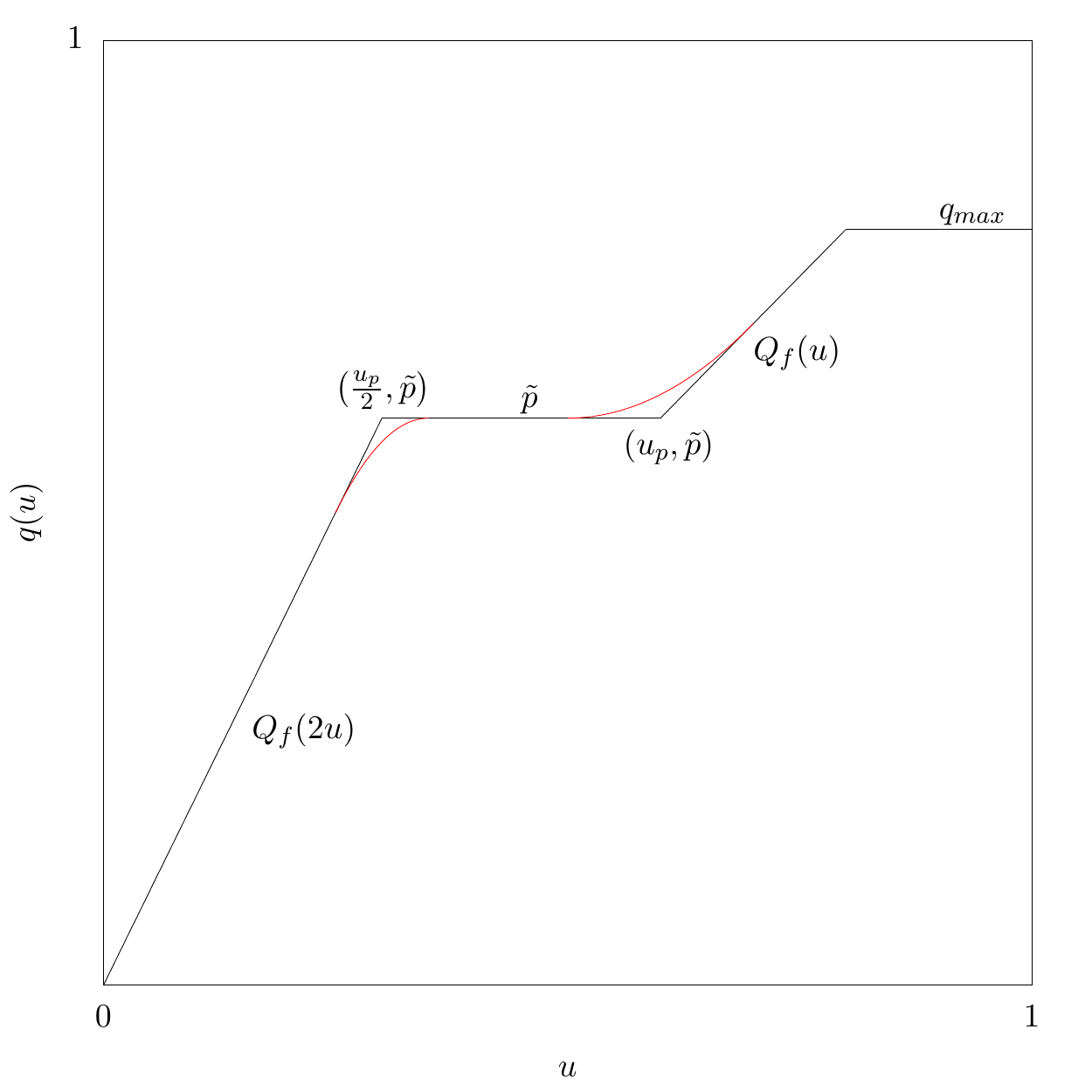}
\caption{Perturbation of the same real-replicas overlap $q(u)$.}
\end{figure}

\begin{figure}[ht]		
\centering
		\includegraphics[scale=0.8]{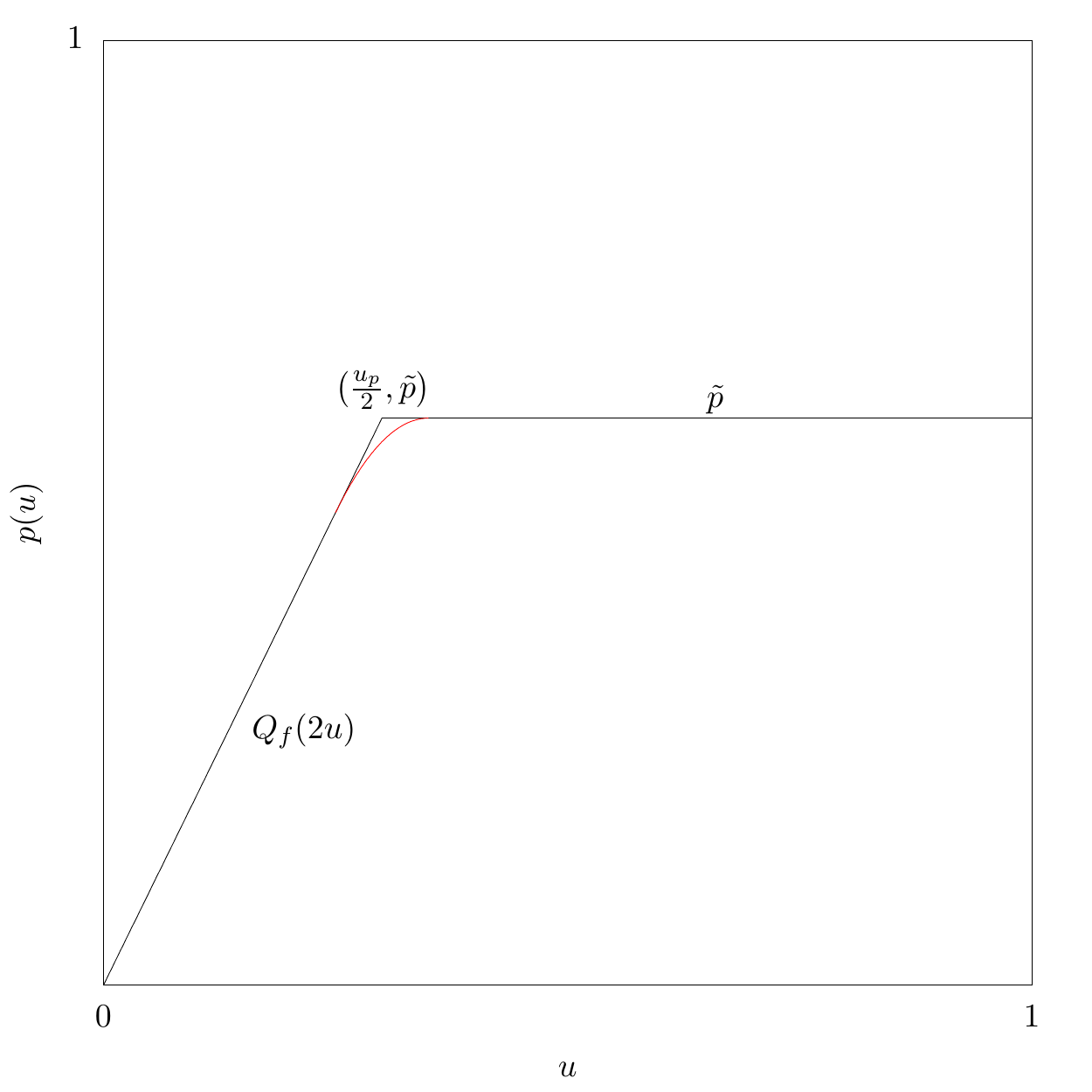}
\caption{Perturbation of the different real-replicas overlap $p(u)$.}
\end{figure}
The variational parameters in the above formula are $\delta$ and $\delta'$, the intervals over which we interpolate the unperturbed solution with the second order polynomial. The constraint imposed on the solution is to have different values of $\tilde{p}(x)$ at the boundaries of the system. This in turn forces a non-vanishing derivative $\frac{d \tilde{p}}{dx}$ at least in some regions of the block. This derivative takes the place of $|p_1 - p_2|/L$ as the perturbative parameter.

The free energy density variation as a function of $\delta$ and $\delta'$ takes the form:
\begin{equation}
\delta f = \frac{31 \delta^5}{10260 y^3} - \frac{\delta^4 \tilde{p}}{324 y^2} + \frac{\chi \delta}{9 y} + \frac{\chi \delta'}{18 y} - \frac{\delta^3 \delta'^2}{9720 y^3} + \frac{\delta^2 \delta'^2 \tilde{p}}{648 y^2} - \frac{\tilde{p} \delta'^4}{5184 y^2} - \frac{11 \delta'^5}{3265920 y^3}
\end{equation}
where it was introduced the variable $\chi = 3 y \left(\frac{d \tilde{p}}{dx} \right)^2$.  
Maximizing over the variational parameters we find the free energy density cost:
\begin{equation}
\delta f = 0.673659\ y^{-\frac{1}{2}} \chi^{\frac{5}{4}} = 0.673659\ y^{\frac{3}{4}}\left(\frac{d \tilde{p}}{dx} \right)^{\frac{5}{2}}
\end{equation}
The solution of this variational problem is obtained for parameters:
\begin{equation}
\delta = \frac{\delta'}{2} = 3.78933\ y^{\frac{1}{2}} \chi^{\frac{1}{4}} 
\end{equation}
Finally maximizing the free energy density variation with respect to the functional form of $\tilde{p}(x)$ we obtain a linear variation of the overlap (and thus a constant free energy density) over the whole block:
\begin{equation}
\tilde{p}(x) = p_1 \left(1- \frac{x}{L}\right) + p_2 \frac{x}{L} \qquad \frac{d \tilde{p}}{d x} = \frac{p_2 - p_1}{L}
\end{equation} 

\begin{equation}
\delta f = 0.673659\ y^{\frac{3}{4}}\left(\frac{p_2 - p_1}{L} \right)^{\frac{5}{2}}
\end{equation}
With this free energy density cost we finally obtain the free energy variation over a volume $V = L^d$:
\begin{equation}
\delta F = 0.673659\ y^{\frac{3}{4}}\left|p_2 - p_1\right|^{\frac{5}{2}}L^{d - \frac{5}{2}}
\end{equation}
and we see that the free energy cost does not grow with the volume for dimensions $d \leq \frac{5}{2}$, indicating that the fluctuations destroy replica symmetry breaking under the critical dimension $D_{LC} = \frac{5}{2}$. 

\section{New results}

\subsection{Higher order continuity}
\label{highorder}
We modified the original calculation to check its stability changing the function space over which the free energy is maximized. The maximization process in the original paper was constrained to variations contained in the space of quadratic polynomials. We improved the smoothness of the solution and changed the space of variations to higher degree polynomials by imposing continuity of the first $k$-derivatives at the insertion points of the polynomial. Our family of perturbed solutions takes the form:
\begin{equation}
q_k(x,u) = \begin{cases} q_{\text{min}} \qquad & 0\leq u \leq u_0  \\ \frac{2 u }{3 y} \qquad & u_0 <  u  \leq u_1 \\  p^k_{\delta}(x,u) \qquad & u_1 <  u \leq u_2  \\ \tilde{p}(x) \qquad & u_2 <  u \leq u_3  \\
p^k_{\delta'}(x,u) & u_3 <  u \leq u_4 \\ \frac{u}{3 y} \qquad  & u_4 <  u \leq u_5  \\ q_{\text{max}} \qquad & 	u_5 <  u \leq 1 \end{cases}
\end{equation}
\begin{equation}
p_k(x,u) = \begin{cases} q_{\text{min}} \qquad & 0\leq u \leq u_0  \\ \frac{2 u }{3 y} \qquad & u_0 <  u  \leq u_1 \\  p^k_{\delta}(x,u) \qquad & u_1 <  u \leq u_2  \\ \tilde{p}(x) \qquad & u_2 <  u \leq 1   \end{cases}
\end{equation}
\begin{equation}
p_1 \leq \tilde{p}(x) \leq p_2
\end{equation}
Now $p^k_{\delta}(x,u)$ and $p^k_{\delta'}(x,u)$ are the lowest degree polynomials in $u$ ensuring the continuity of the solutions up to the $k$-th order derivative. In the original solution interpolating with second order polynomials we had a discontinuity of the second derivative at the insertion points; here we can impose continuity to arbitrary order. To check the stability of the solution we variated the solution imposing continuity of second order derivatives, obtaining for the free energy density cost the result:
\begin{equation}
\delta f_{2} = 0.673328\ y^{\frac{3}{4}}\left(\frac{p_2 - p_1}{L} \right)^{\frac{5}{2}}
\end{equation}
As we see the functional form of the expression is completely unchanged, the only difference being a small reduction in the numerical coefficient. Finally evaluating the free energy cost on a solution with continuous third order derivatives at the insertion points we obtain:
\begin{equation}
\delta f_{3} = 0.671176\ y^{\frac{3}{4}}\left(\frac{p_2 - p_1}{L} \right)^{\frac{5}{2}}
\end{equation}
It can be noted that with continuity of the second and third order derivatives of the solution at the insertion points the change in the free energy cost is \emph{negative}, so we are slightly moving away from the real maximum\footnote{We checked the result also for linear polynomials, obtaining the free energy density cost $\delta f_{0} = 0.549271 \ y^{\frac{3}{4}}\left(\frac{p_2 - p_1}{L} \right)^{\frac{5}{2}}$, the same functional form but lower coefficient than for the second order polynomials, as expected.}.

\subsection{Extended model with local perturbations}
\label{pol6}

After a careful review of the effect of different boundary conditions on the free energy \eqref{truncated} we want to generalize the result to different systems. The saddle point associated to the previous free energy is a solution linear in replica space, which thus implies a constant probability for the overlap. this fact makes it impossible to study the dependence of the free energy cost on the overlap. In \cite{Franz2009} it was shown that in a hierarchical spin-glass model the same calculation as above can be done, bringing for small $\left|p_2 - p_1\right|$ the free energy density cost:
\begin{equation}
\delta f_{\text{hier}} \propto P\left(\overline{p}\right)\left|p_2  - p_1\right|^3
\end{equation}
where $P(q)$ is the overlap probability and $\overline{p}=\frac{p_1 + p_2}{2}$ is the average overlap.

The free energy density obtained in the previous section trivially satisfies a similar relation (though with different exponents), but given that the probability $P(q)$ is itself a constant it is difficult to establish the significance of this result. 
We want to consider an extension of the truncated model which generates a non-linear solution $q(u)$, and as a consequence a non-trivial overlap probability $P(q)$.
In this section we extend the model by adding to the free energy density a generic polynomial function local in replica space; relabeling the free energy \eqref{truncated} $F_0$ we want to study the system specified by:
\begin{equation}
-2 n F_{\text{loc}} = -2 n F_0 + 2 z  \int d^D x \sum_{ab} g\left(Q_{ab}(x) \right) 
\end{equation} 
with 
\begin{equation}
g(q) = g_6  q^6 + g_8  q^8 + g_{10} q^{10} + ...
\end{equation}
and only even terms are included in order to preserve the $\mathbb{Z}_2$ symmetry of the original Hamiltonian.
With ``local in replica space" we mean that there are no terms proportional to $z$ which couple overlap matrix elements $Q_{ab}$ for different values of the codistance between replica indices $ab$ (in the continuum limit this would translate to a different codistance index $u$). This peculiarity of the new ``interaction terms" allows us to extend to this free energy many of the results derived in the FPV calculation.

As a first step it is easy to show that the saddle point equations derived from such a free energy (with only one real replica) are algebraic and local in the matrix elements $Q_{ab}$, as in the truncated model. This implies that the solution cannot depend on the temperature and magnetic field for any $u \in (u_0, u_1)$, where $q(u_0) = q_{\text{min}}$ and $q(u_1) = q_{\text{max}}$. 
This directly implies that there can be no dependence of the free energy cost on the temperature and magnetic field, for the same argument given in section \ref{trunc model section}: no term in $F_0$ is temperature or magnetic field-dependent, and given that all the terms in $g\left(Q_{ab}\right)$ have only one set of replica indices their variations have the form:
\begin{equation}
\delta g_m(Q_{ab}) \propto  Q_{ab}^{m-j}\delta Q_{ab}^{j} 
\end{equation}
They are different from zero only when $\delta Q_{ab}$ is different from zero, where there is no dependence on temperature and magnetic field in $Q_{ab}$.

Given the algebraic nature of the saddle point equations they are easy to solve for any polynomial $g(Q_{ab})$. As an example the exact solution for the lowest order non-trivial polynomial, $g(Q_{ab}) = \frac{z}{6!}Q_{ab}^6$, has the form: 
\begin{equation}
Q_{ab}(x) \to q\left(x , u\right) = q\left( u\right) = \begin{cases} q_{\text{min}}\qquad u \leq u_0  \\  \tilde{q}(u) \qquad u_0 \leq u \leq u_1  \\ q_{\text{max}}\qquad u \geq u_1 \end{cases}
\end{equation}
\begin{equation}
\tilde{q}(u) = \frac{1.44 \left(\sqrt{z} w u + \sqrt{z w^2 u^2 + 24 y^3}\right)^{2/3}- 4.16 y}{\left(z^2 w u + z^{3/2}\sqrt{z w^2 u^2 + 24 y ^3}\right)^{1/3}}
\end{equation}
The last expression as a power series in $u$ becomes:
\begin{equation}
\tilde{q}(u) = \frac{u}{3 y} - \frac{z u^3}{486 y^4} + \frac{z^2 u^5}{26244 y^7} + O(u^7)
\end{equation}
The inverse function $u(q)$ and the associated bulk overlap probability have the simpler form:
\begin{equation}
u(q) = 3 y q + \frac{z}{12} q^3 \qquad P(q) = \dot{u}(q) = 3 y + \frac{z}{4} q^2
\end{equation}
Finally we can evaluate the free energy cost with this extended free energy using the same variational approach as in the previous section. We can simplify the calculation performing it perturbatively in the coupling constant $z$; at the first perturbative order we obtain:
\begin{equation}
\delta f_{Q^6}(x) = 0.673659 \ y^{\frac{3}{4}}\left(\frac{d \tilde{p}}{dx} \right)^{\frac{5}{2}} \left(1 + \frac{z}{16 y}\tilde{p}^2(x) + O\left(z^2, L^{-1}\right)\right)
\end{equation}
We see that at this order of approximation the free energy density cost is indeed proportional to $P(q)^{\frac{3}{4}}$. We verified this proportionality for polynomials of order up to 24, and derived it analytically in section \ref{anres}.

The last step is to maximize the free energy over the functional form of $\tilde{p}(x)$. We find a functional form different from the linear one of FPV; in particular we obtain:
\begin{equation}
\frac{d \tilde{p}(x)}{dx} = \gamma \left( 1 - \frac{z}{40 y} \tilde{p}^2\right) + O\left(z^2\right), \quad \gamma = \frac{p_2 - p_1}{L} + \frac{z}{120 y}\frac{p_2^3 - p_1^3}{L}
\end{equation}
This functional form implies a homogeneous distribution of the free energy cost over the block:
%\begin{equation}
%\delta f_{Q^6} = 0.673659\ y^{\frac{3}{4}}\left(\frac{p_2 - p_1}{L} \right)^{\frac{5}{2}}\left(1 + \frac{z}{48 y} \frac{p_2^3 - p_1^3}{p_2 - p_1} + O\left(z^2, L^{-1}\right)\right)
%\end{equation} 
\begin{equation}
\delta f_{Q^6} = 0.673659\ y^{\frac{3}{4}}\left(\frac{p_2 - p_1}{L} \right)^{\frac{5}{2}}\left[1 + \frac{z}{48 y} \left(p_2^2 + p_1 p_2 + p_1^2 \right) + O\left(z^2, L^{-1}\right)\right]
\end{equation} 
%Qui pensavo di mettere la lagrangiana estesa con i polinomi di tipo $\sum_{ab} Q^n_{ab}$, calcolando il costo perturbativamente ma facendo vedere non perturbativamente che il costo non dipende da temperatura e campo magnetico.
\subsection{Extended model for the Edwards-Anderson free energy}
\label{sectionEA}
In this section we want to perform the calculation for a system with a free energy closer to the real Edwards-Anderson one. We can obtain the truncated model of section \ref{trunc model section} expanding the Edwards-Anderson effective Hamiltonian up to order four in powers of the field, and retaining only the fourth order term responsible for the replica symmetry breaking. This is the minimal model to describe a system with replica symmetry breaking, and the solution for it is the linear overlap function we met in section \ref{trunc model section}. If we want to study what happens when the solution of the system becomes non-linear it is sufficient to add the remaining fourth order terms to the free energy. 
The free energy including all the fourth order terms has the form\footnote{In the Edwards-Anderson effective Hamiltonian the coupling constant are fixed to values of order one, but in most of this section we keep them as free parameters.}:
\begin{multline}
\label{extended}
-2 n F = \int d^D x \Bigg[- \texttt{Tr}(|\nabla Q (x)|^2)  + \tau \texttt{Tr} Q^2(x) + \frac{1}{3} 
\texttt{Tr} Q^3(x) +  \\ + \frac{y}{4}\sum_{ab}Q_{ab}(x)^4 + \frac{z}{12} \texttt{Tr}Q^4(x)  - \frac{t}{2}\sum_{abc}Q^2_{ab}(x)Q^2_{bc}(x)  +  h^2 \sum_{ab} Q_{ab}(x) \Bigg]
\end{multline}
We call the two additional terms respectively $F_z$ and $F_t$ (denoting as in the previous sections densities with lower case letters). Their contribution to the saddle point equations is given by:
\begin{equation}
- n \frac{\delta F_z}{\delta Q_{ab}(x)}\Bigg|_{\delta Q_{ab}=0}  = \frac{z}{6}Q_{ab}^3(x) 
\end{equation}
\begin{equation}
- n \frac{\delta F_t}{\delta Q_{ab}(x)}\Bigg|_{\delta Q_{ab}=0}  = -  \frac{t}{n} \ Q_{ab}(x) \texttt{Tr} Q^2(x) 
\end{equation}
The new saddle point equations for a solution homogeneous in space are thus:
\begin{equation}
\left(\tau - \frac{t}{n} \texttt{Tr} Q^2 \right) Q_{ab} + \frac{\left(Q^2\right)_{ab}}{2} + \frac{y}{2} Q_{ab}^3 + \frac{z}{6} \left(Q^3\right)_{ab} + h^2 = 0 
\end{equation}
The only contribution of the $F_t$ term is to effectively increase (or reduce, for negative $t$) the reduced temperature of the system. This can be seen by the fact that in the saddle point equations the reduced temperature appears only in the combination $\tau' = \left(\tau - \frac{t}{n} \texttt{Tr} Q^2(x) \right)$. The term $F_z$, on the other hand, deforms the functional form of the solution making it non linear (and giving it an explicit dependence on the \emph{effective} temperature $\tau'$).
Solving the equation with $z=0$ we would have the same linear solution as section \ref{trunc model section}, dependent on the temperature only in $q(u) = q_{\text{max}}$ and on the magnetic field only in $q(u) = q_{\text{min}}$. The value of $q_{\text{max}}$ however would depend on the effective temperature $\tau'$ instead of $\tau$; $\tau'$ has a dependence both on the temperature and magnetic field through the term $\texttt{Tr}Q^2$, but given that $q(u) \leq q_{\text{max}} \approx \tau'$ the corrections will always be smaller than $\tau$ if $t$ is of order one.   

Solving the saddle point equations with the $F_z$ term the situation changes more radically; the solution becomes non linear and acquires a dependence on the effective temperature for every value of the overlap $q(u)$. We have the explicit form:
\begin{equation}
\tilde{q}(u) = \frac{u}{3 y} \sqrt{\frac{1 - 2 z \tau'}{1+z \frac{u^2}{3 y}}} 
\end{equation}
We want now to study the dependence of the free energy cost on the temperature and magnetic field. Given the explicit dependence of the solution of the saddle point equations on the effective temperature $\tau'$ we expect a similar dependence also in the free energy cost. We already saw that the free energy cost of the reduced model $F_0$ cannot depend explicitly on the temperature or the magnetic field; the only dependence can be trough $q(u)$, and thus trough $\tau'$. We can check that the same happens for the two additional terms $F_z$ and $F_t$. On the saddle point solution we have:
\begin{equation}
- n \delta f_z  = \frac{z}{12}\texttt{Tr}\left(4 Q \delta Q^3 + 6 Q^2 \delta	Q^2 + \delta Q^4  \right)
\end{equation}
\begin{equation}
- n \delta f_t  = -  \frac{t}{2} \sum_{abc}\left( 2  Q_{ab}^2 \delta Q_{bc}^2 + 4 Q_{ab}\delta Q_{ab}\delta Q_{bc}^2 + \delta Q_{ab}^2 \delta Q_{bc}^2 	\right)
\end{equation}
Taking the derivative of the right hand side of the the first equation with respect to $Q_{ab}$ we obtain:
\begin{equation}
- n \frac{\partial \delta f_z}{\partial Q_{ab}} = z \left[\frac{\left(\delta Q^3 \right)_{ab}}{3} + \left(Q \delta Q^2 \right)_{ab} \right]
\end{equation}
It is easy to show that both terms vanish when $ab$ is such that\footnote{We use the fact that $\texttt{Tr} \delta Q = 0$ from the equation for $q_{\text{max}}$ evaluated in $ab =0$.} $Q_{ab}=q_{\text{min}}$. The same is not true when $ab$ is such that $Q_{ab}=q_{\text{max}}$, but the only temperature entering the equation for $q_{\text{max}}$ is the \emph{effective} temperature, so the dependence is on $\tau'$.
As for the terms in $\delta f_t$ they can be rewritten as:
\begin{equation}
- n \delta f_t  = -  \frac{t}{2 n} \left( 2 \texttt{Tr} Q^2 \texttt{Tr}\delta Q^2 + 4 \texttt{Tr}Q\delta Q \texttt{Tr}\delta Q^2 + \left(\texttt{Tr}\delta Q^2 \right)^2 	\right)
\end{equation}
and we already showed that for terms of this kind there is no coupling between values of the overlap for different values of the index $ab$. As a consequence there can be no dependence on $q_{\text{min}}$ or $q_{\text{max}}$ when $\delta Q_{ab}$ is different from zero - the only possible dependence is trough $q(u)$.

The previously examined terms thus cannot depend explicitly on the temperature and magnetic field; the only dependence on these parameters is trough $\tau'$. In addition we expect an explicit dependence on $\tau'$, not only trough $q(u)$, only from the term $\delta f_z$.

From the saddle point solution we obtain the overlap probability:
\begin{equation}
P\left(q\right) =  \frac{3 y\left(1 - 2 z \tau'\right)}{\left( 1 - 3y z q^2 - 2 z \tau' \right)^{3/2}}
\end{equation}
We proceed to evaluate the free energy cost perturbing the reduced model at the first order in the coupling constant $z$. The correction given by the term $f_t$ to the temperature can be found solving a self-consistency equation for $\texttt{Tr} Q^2$, and all other contributions proportional to $t$ vanish. 
We obtain the free energy cost:
\begin{equation}
\delta f_{\texttt{Tr} Q^4}(x) =  0.673659 \ y^{\frac{3}{4}}\left(\frac{d \tilde{p}}{dx} \right)^{\frac{5}{2}} \left(1 + \frac{15 z y}{4} \tilde{p}^2(x) + z \tau' + O\left(z^2, L^{-1}\right)\right)
\end{equation}
The overlap probability expanded at this order of approximation in $z$ is:
\begin{equation}
P\left( q \right)=3 y \left(1 + \frac{9 z y }{2} \tilde{p}^2(x) + z \tau' + O\left(z^2 \right)\right)
\end{equation}
We can see that in this case the free energy cost is not proportional to any power of $P(q)$; we have the same functional dependence on $\tilde{p}$ and the effective temperature, but slightly different numerical coefficients. The other difference with the previously analyzed cases is the dependence on the effective temperature that we already saw. These differences can be traced back to the different nature of the saddle point equations: while in the previous cases they reduced to ``local", algebraic equations, in this case they are differential equations, the solutions of which depend on boundary conditions. We note also that the only effect of $f_t$ has been to modify the temperature, all the other terms being subleading in $L^{-1}$. 

Having recovered the free energy density cost as a function of the overlap and temperature we can maximize over the functional form of the overlap, to find the free energy as a function of the temperature:
%\begin{equation}
%\delta f_{\texttt{Tr} Q^4} =  0.673659 \ y^{\frac{3}{4}}\left(\frac{p_2 - p_1}{L} \right)^{\frac{5}{2}} \left[1 + z \tau' + \frac{5}{4}z y \left(\frac{p_2^3 - p_1^3}{p_2 - p_1}\right)  + O\left(z^2, L^{-1}\right)\right]
%\end{equation}
\begin{equation}
\delta f_{\texttt{Tr} Q^4} =  0.673659 \ y^{\frac{3}{4}}\left(\frac{p_2 - p_1}{L} \right)^{\frac{5}{2}} \left[1 + z \tau' + \frac{5}{4}z y \left(p_1^2 + p_1 p_2 + p_2^2\right)  + O\left(z^2, L^{-1}\right)\right]
\end{equation}
This result can be compared to the one recently reported in \cite{Maiorano18}. There the authors study the variation of the interface \emph{energy} as a function of the number of dimensions. At the order of approximation of the calculation in \cite{Franz94} the energy of the interface would be zero, being no dependence of the free energy on the temperature. At lowest non trivial order we find for the internal energy the value:
\begin{equation}
\delta E_{TrQ^4} = 0.673659 \ y^{\frac{3}{4}} z \left(p_2  - p_1 \right)^{5/2} L^{D-\frac{5}{2}} + O\left( z^2, \tau, L^{D-\frac{7}{2}} \right)
\end{equation} 
This value is in agreement with the findings of \cite{Maiorano18}.

The results of this section can be confirmed adopting a different approximation scheme. The reduced model on which the calculations are based is valid only for a system close to the critical temperature. We can then limit the perturbations to the same regime, and approximate the free energy cost using the temperature as small parameter. Using this approximation scheme the results are valid for all values of the coupling constants, so we can substitute their values to obtain the Edwards-Anderson model (in which they are of order unity). The free energy density cost in this approximation (here up to second order in the temperature) is:
\begin{equation}
\delta f_{\texttt{Tr} Q^4} =  0.673659 \ y^{\frac{3}{4}}\left(\frac{p_2 - p_1}{L} \right)^{\frac{5}{2}} \left[1 + z \tau' + \frac{5}{4} z y \left(\frac{p_2^3 - p_1^3}{p_2 - p_1}\right) + \frac{3}{2} z^2 \tau'^2 + O\left(\tau^3, L^{-1}\right)\right]
\end{equation}
As we can see the result is perfectly equivalent to the previous one at the first order in $z$, higher orders in $z$ being connected to higher orders in $\tau$.
The last expression allows us also to evaluate the variation of the internal energy as a function of the temperature:
\begin{equation}
\delta E_{TrQ^4} = 0.673659 \ y^{\frac{3}{4}} z \left(p_2  - p_1 \right)^{5/2} L^{D-\frac{5}{2}}\kappa(\tau) + O\left(\tau^2, L^{D-\frac{7}{2}} \right)
\end{equation}
with $\kappa(\tau) = \frac{d}{d \tau}\left( 1 + z \tau' + \frac{3}{2} z^2 \tau'^2  \right)$.
We obtain the exact value for the Edwards-Anderson model (at this order of approximation in the temperature) substituting the values $z = 3$, $y=2/3$.

Finally we note that the scaling $L^{D - \frac{5}{2}}$ is stable in all approximation schemes and valid for any value of the ovelap difference $|p_2 - p_1|$. In particular the scaling shows no singularities when the overlap goes to zero. 
%Questa dovrebbe essere la parte in cui vediamo gli effetti di $z \sum_{abc} Q_{ab}^2 Q_{bc}^2 + u \texttt{Tr} Q^4$, calcoliamo il costo perturbativamente ma dimostriamo la dipendenza del costo energetico dalla temperatura e dal campo magnetico in modo non perturbativo.

\section{Analytic results and general models}
\label{anres}
In this section we show a number of analytic results generalizing the calculations of the previous sections. The analytic derivation allows us to consider what happens in a general model having the same structure of the two previously analyzed and in which the same variational procedure is applied. In addition the model described in this section describes accurately the Edwards-Anderson free energy up to order six in the overlap field.
This general free energy can be written as: 

\begin{equation}
\mathcal{L}(x)= \Bigg[- \frac{1}{2} \texttt{Tr}(|\nabla Q (x)|^2)  + \sum_{j=1}^{\infty}\sum_{i=2}^{\infty} t_{ij} \left(\texttt{Tr} Q^i(x)\right)^j + \sum_{k=1}^{\infty} y_k \sum_{ab} Q_{ab}(x)^k \Bigg]
\end{equation}
the models previously analyzed can be recovered from the general one keeping different from zero only the coupling constants:
\begin{equation}
\left(t_{21}, t_{31}\right) = \left(\frac{\tau}{2}, \frac{1}{6}\right), \quad \left(y_{1},y_{4},y_{6}\right)=\left(\frac{h^2}{2},\frac{y}{8},\frac{z}{6!}\right) 
\end{equation}
for the sixth-order polynomial model and:
\begin{equation}
\left(t_{21},t_{31},t_{41}, t_{22}\right)=\left(\frac{\tau}{2}, \frac{1}{6}, \frac{z}{4!}, -\frac{t}{4 n}\right), \quad \left(y_{1},y_{4}\right)=\left(\frac{h^2}{2},\frac{y}{8}\right) 
\end{equation}
for the Edwards-Anderson model\footnote{It can be easily proved that the term $\sum_{abc}Q^2_{ab}Q^2_{bc}$ is proportional to $\left(\texttt{Tr}Q^2\right)^2$.}. 

%As a first step we will analyze the model with only a local potential:
%\begin{equation}
%\mathcal{L}(x)= \Bigg[- \frac{1}{2} \texttt{Tr}(|\nabla Q (x)|^2)  + \frac{\tau}{2}\texttt{Tr} Q^2(x) + \frac{1}{6}\texttt{Tr} Q^3(x) + V\left(Q(x)\right)  \Bigg]
%\end{equation} 
%where $V\left(Q\right) = \sum_{k=1}^{\infty} y_k \sum_{ab} Q_{ab}(x)^k$.

To solve the general saddle point when there are no constraints on the boundary conditions we use the standard methods, but here we outline the different steps because they will be useful in exploiting all the symmetries of the theory. For the saddle point we have:
\begin{equation}
\sum_{j=1}^{\infty}\sum_{i=2}^{\infty} i\ j\ t_{ij} \left(\texttt{Tr} Q^i\right)^{j-1} \left(Q^{i-1}\right)_{ab} + \sum_{k=1}^{\infty} k\ y_k  Q_{ab}^{k-1} = 0
\end{equation}
We can see that the terms proportional to $t_{ij}$ with $j>1$ give the only contribution of rescaling the coupling constants of $Q^{i-1}$. This is a generalization of what we saw in the case of $\sum_{abc}Q^2_{ab}Q^2_{bc}$, and in the same way we can consider effective coupling constants to absorb these contributions.
Using this strategy the last equation becomes:
\begin{equation}
\sum_{i=2}^{\infty} i\ t'_i \left(Q^{i-1}\right)_{ab} + \sum_{k=1}^{\infty} k\ y_k  Q_{ab}^{k-1} = 0
\end{equation}
\begin{equation}
t'_i = \sum_{j=1}^{\infty} j\ t_{ij} \left(\texttt{Tr} Q^i\right)^{j-1}
\end{equation}
Now taking the continuum limit and deriving with respect to $Q_{u}$ we obtain:
\begin{equation}
\label{eigenvalueeq}
\sum_{i=2}^{\infty} i(i-1) t'_i \ \Lambda_{Q_{u}}^{i-2} + \sum_{k=2}^{\infty} k(k-1) y_k\  Q_{u}^{k-2} = 0
\end{equation}
$\Lambda_{Q_{u}}$ being the eigenvalues of the matrix $Q_{ab}$ in the continuum limit\footnote{A short review of hierarchical matrices diagonalization in the continuum limit is given in the appendix.}. Deriving again:
\begin{equation}
- u \sum_{i=3}^{\infty} \frac{i!}{(i-3)!} t'_i \ \Lambda_{Q_{u}}^{i-3} + \sum_{k=3}^{\infty} \frac{k!}{(k-3)!} y_k\  Q_{u}^{k-3} = 0
\end{equation}
From the last formula we can obtain the replica coordinate $u$ as a functional of $Q_u$ ($\Lambda_{Q_{u}}$ is not a function of $Q_u$):
\begin{equation}
\label{uqf}
u = \frac{\sum_{k=3}^{\infty} \frac{k!}{(k-3)!} y_k\  Q_{u}^{k-3}}{ \sum_{i=3}^{\infty} \frac{i!}{(i-3)!} t'_i \  \Lambda_{Q_{u}}^{i-3}} 
\end{equation}
The last equation becomes a definition of the function $u(Q)$ if $t'_i=0$ for every $i>3$. In this case except for the term $\texttt{Tr} Q^3$ we have only local interactions in the effective Hamiltonian, i.e. we don't have couplings of matrices $Q_u$ with different replica index. This is obviously the case of polynomial interactions we analyzed above, and we will see why in this case the free energy cost is proportional to a power of the overlap probability $P(Q)$. For now we can notice that when this simplification occurs we have:
\begin{equation}
P(Q) = \dot{u}(Q) =  \frac{1}{6\ t'_3} \sum_{k=4}^{\infty}  \frac{k!}{(k-4)!} y_k\  Q^{k-4}
\end{equation}
When we have some non-local terms on the other hand we have to solve equation \eqref{eigenvalueeq} to find $\Lambda_{Q_u}$ as a function of $Q_u$, but even when the equation can be solved $\Lambda_{Q_u}$ will depend also on the parameters $t'_i$, among which we find for example the temperature (but not the magnetic field, which can appear only trough the $t'_is$). 

For the Edwards-Anderson model we have from \eqref{uqf}:
\begin{equation}
\label{uqfEA}
u = \frac{3 y Q}{1 + z \Lambda_{Q_u}}
\end{equation} 
and from \eqref{eigenvalueeq}:
\begin{equation}
2 \tau' + 2 \Lambda_{Q_u} + z \Lambda_{Q_u}^2 + 3 y Q_u^2 = 0
\end{equation}
The last equation gives us:
\begin{equation}
\Lambda_{Q_u} = -\frac{1}{z} + \sqrt{\frac{1}{z^2} - \frac{3 y Q_u^2 + 2 \tau'}{z}} = -\frac{1}{z} + \sqrt{\frac{1}{z^2} - \frac{\left(\Lambda_{ Q_u}\right)_{z=0}}{z}}
\end{equation}
%\begin{equation}
%\Lambda_{Q_u} = -\frac{1}{z} + \sqrt{\frac{1}{z^2} - \frac{3 y Q_u^2 + 2 \tau'}{z}} \xrightarrow[z \to 0]{} - \left( \frac{3 y Q_u^2}{2} +  \tau'\right)
%\end{equation}
Substituting this expression in \eqref{uqfEA} we readily obtain, to the first order in $z$, the same results we saw in section \ref{sectionEA}:
\begin{equation}
u(Q) \approx 3 y Q \left( 1 - \frac{z}{2} \left(\Lambda_{ Q_u}\right)_{z=0}\right) = 3 y Q \left( 1 +  z \frac{3 y Q^2}{2} + z \tau' \right)
\end{equation}
\begin{equation}
P(Q) \approx 3 y \left( 1 +   \frac{9 y z}{2}Q^2 + z \tau' \right)
\end{equation}
If we now impose boundary conditions $p_1 = p_2 = \tilde{p}$ to a set of coupled real replicas of this general model the solution will have the same structure we saw in the FPV calculation \cite{Franz2005}:
\begin{equation}
Q_{ab}(x) \to q_0(x,u) = q_0(u) = \begin{cases} q_{\text{min}} \qquad & 0\leq u \leq \frac{u_0}{2}  \\ Q_f (2 u) \qquad & \frac{u_0}{2} <  u  \leq \frac{u_p}{2} \\ \tilde{p} \qquad & \frac{u_p}{2} <  u \leq u_p  \\ Q_f(u) \qquad  & u_p <  u \leq u_1  \\ q_{\text{max}} \qquad & 	u_1 <  u \leq 1 \end{cases}
\end{equation}
\begin{equation}
P_{ab}(x) \to p_0(x,u) = p_0(u) = \begin{cases} q_{\text{min}} \qquad & 0\leq u \leq \frac{u_0}{2} \\ Q_f(2 u) \qquad & \frac{u_0}{2} <  u  \leq \frac{u_p}{2} \\ \tilde{p} \qquad & \frac{u_p}{2} <  u \leq 1  \end{cases}
\end{equation}
\begin{equation}
\tilde{p}(x) = \tilde{p} = p_1 = p_2
\end{equation}
where $Q_f(u)$ is the inverse of the function $u(Q)$ we derived in the beginning of the section, $u_p = u\left(\tilde{p} \right)$ and the rest of the parameters have the same meaning as in the FPV calculation.  

The solution of the problem with equal boundary conditions can be used as a basis for a variational problem when $p_1 \neq p_2$, and as in the rest of the paper we assume the variations to be localized in two small neighborhoods of the breaking points in replica space.   
\begin{equation}
q(x,u) = \begin{cases} q_{\text{min}} \qquad & 0\leq u \leq u_0  \\ Q_f (2 u) \qquad & u_0 <  u  \leq u_1 \\  P_{\delta}(x,u) \qquad & u_1 <  u \leq u_2  \\ \tilde{p}(x) \qquad & u_2 <  u \leq u_3  \\
P_{\delta'}(x,u) & u_3 <  u \leq u_4 \\ Q_f (u) \qquad  & u_4 <  u \leq u_5  \\ q_{\text{max}} \qquad & 	u_5 <  u \leq 1 \end{cases}
\end{equation}
\begin{equation}
p(x,u) = \begin{cases} q_{\text{min}} \qquad & 0\leq u \leq u_0  \\ Q_f (2 u) \qquad & u_0 <  u  \leq u_1 \\  P_{\delta}(x,u) \qquad & u_1 <  u \leq u_2  \\ \tilde{p}(x) \qquad & u_2 <  u \leq 1   \end{cases}
\end{equation}
\begin{equation}
p_1 \leq \tilde{p}(x) \leq p_2
\end{equation}
where $P_{\delta}$, $P_{\delta'}$ are interpolating polynomials, the solution has continuous first derivatives at the breaking points and again the parameters have the same meaning of the FPV ones.
In the FPV calculation and in sections \ref{pol6}, \ref{sectionEA} the two variational parameters $\delta$, $\delta'$ were varied independently, and the maximum of the free energy was found for $\delta' = 2 \delta$. This can be shown to be a consequence of the ultrametric structure of the solution, and in addition it implies the identities:
\begin{equation}
\label{simplif}
\int_0^1 \left[ \delta q(x,u)^{(2 m + 1)}  + \delta p(x,u)^{(2 m + 1)}  \right] du = 0 \qquad \forall\ m \in \mathbb{N}
\end{equation}
with $\delta q(x,u) = \left[q(x,u)- q_0(u)\right]$, $\delta p(x,u) = \left[p(x,u)- p_0(u)\right]$ are functions localized around the breaking points $\frac{u_p}{2}$, $u_p$. 

We want now to derive analytically the free energy density cost for the variational problem associated to this model. The free energy to be maximized can be written as:
\begin{equation}
- n F = \int dx \Bigg[- \frac{1}{2}\texttt{Tr}|\nabla Q (x)|^2  + \texttt{Tr}A(Q(x)) + \sum_{ab} B\left(Q_{ab}(x)\right) \Bigg]
\end{equation}
where
\begin{equation}
A(Q)_{ab} = \sum_{i=1}^{\infty} t'_i \left(Q^i\right)_{ab} \qquad B(Q_{ab}) =  \sum_{k=1}^{\infty} y_k\ Q_{ab}^k 
\end{equation}
The variations of the different terms are:
\begin{equation}
\label{kinetic}
\frac{1}{2} \delta \texttt{Tr}|\nabla Q (x)|^2 = \frac{1}{2}\texttt{Tr}|\nabla \delta Q (x)|^2 + \texttt{Tr}\nabla \delta Q (x)\nabla Q (x)
\end{equation}
\begin{equation}
\label{trace}
\delta \texttt{Tr}A(Q(x)) = \sum_{k=2}^{\infty} \frac{1}{k!}\texttt{Tr}\left(A^{(k)}(Q(x))\delta Q^k(x)\right)
\end{equation}
\begin{equation}
\label{local}
\sum_{ab} \delta B\left(Q_{ab}(x)\right) = \sum_{k=2}^{\infty}\frac{1}{k!} \sum_{ab} B^{(k)}\left(Q_{ab}(x)\right)\delta Q_{ab}^k(x)
\end{equation}
Assuming that $P_{\delta}$ and $P_{\delta'}$ are polynomials with the same derivative as $Q_f(u)$ at the breaking points, and given the cancellations implied by equation \eqref{simplif} it is easy to derive the contribution of the kinetic and local terms \eqref{kinetic}, \eqref{local}:
\begin{equation}
\delta \texttt{Tr}|\nabla Q (x)|^2 =  \alpha \left(\frac{d \tilde{p}}{dx} \right)^2 \delta
\end{equation}
\begin{equation}
\sum_{ab} \delta B\left(Q_{ab}(x)\right) = \frac{1}{2} B^{(2)}\left( \tilde{p}\right)\texttt{Tr}\left(\delta Q^2\right)  + \beta \frac{ B^{(4)}\left( \tilde{p}\right)}{\dot{u}\left(\tilde{p}\right)^4} \delta^5 + O\left( \delta^6\right)
\end{equation}
With $\alpha = \frac{4}{3}$ and $\beta = - \frac{1}{540}$.
The evaluation of the expression \eqref{trace} is more convoluted; using equation \eqref{eigenvalueeq} we can write the first, second and third term\footnote{The derivation can be found in the appendix.} as:
\begin{equation}
\texttt{Tr}\left(A^{(2)}(Q)\delta Q^2\right) =  - B^{(2)}\left( \tilde{p}\right)\texttt{Tr}\left(\delta Q^2\right) + \left( \gamma \frac{B^{(3)}\left(\tilde{p} \right)}{u_p} + \zeta \frac{B^{(4)}\left(\tilde{p} \right)}{\dot{u}\left(\tilde{p} \right)}  \right) \frac{\delta^5}{\dot{u}\left(\tilde{p} \right)^{3}} + O\left(\delta^6 \right) 
\end{equation}
\begin{equation}
\texttt{Tr}\left(A^{(3)}(Q) \delta Q^3\right) = \left( \eta \frac{B^{(3)}\left(\tilde{p} \right)}{u_p} + \theta \frac{u_p\, \sigma\left(u_p \right)}{\dot{u}\left(\tilde{p} \right) }  \right) \frac{\delta^5}{\dot{u}\left(\tilde{p} \right)^{3}} + O\left(\delta^6 \right)
\end{equation}
\begin{equation}
\texttt{Tr}\left(A^{(4)}(Q)\delta Q^4\right) =  \vartheta \frac{u_p\, \sigma\left(u_p \right)}{\dot{u}\left(\tilde{p} \right)} \frac{\delta^5}{\dot{u}\left(\tilde{p} \right)^3}   + O\left(\delta^6 \right)
\end{equation}
Where we defined the function $\sigma(u) =  \left( \frac{d}{dq}  \frac{B^{(3)}\left(q \right)}{u\left( q \right)}\right)_{q=Q_f(u)}$.
%\frac{1}{u\ q'(u)}\left(\frac{B^{(3)}\left(q_u \right)}{u}\right)'
All the higher order terms are negligible if the terms of order $\delta^5$ do not cancel out. 
The value of the numerical coefficients $\gamma$, $\zeta$, $\eta$, $\theta$ and $\vartheta$ is reported in the appedix. 

Summing up all the previous contributions we obtain the expression for the free energy density cost as a function of $\delta$:
\begin{equation}
\delta f = \alpha \left(\frac{d \tilde{p}}{dx} \right)^2 \delta - \left[\Phi \frac{B^{(4)}\left(\tilde{p} \right)}{\dot{u}\left(\tilde{p} \right)} + \Psi \frac{B^{(3)}\left(\tilde{p} \right)}{u_p}  + \Omega \frac{u_p\, \sigma\left(u_p \right)}{\dot{u}\left(\tilde{p} \right) } \right] \frac{\delta^5}{\dot{u}\left(\tilde{p} \right)^{3}} + O\left(\delta^6 \right)
\end{equation}
with $\Phi=\left(\beta + \frac{\zeta}{2}  \right)$, $\Psi=\left(\frac{\gamma}{2} + \frac{\eta}{6} \right)$ and $\Omega = \left( \frac{\theta}{6} + \frac{\vartheta}{24} \right) $.  

When $A^{(4)}\left(Q \right) = 0$ the terms in square brackets collapse to a numerical factor independent of $\tilde{p}$ and the maximization over $\delta$ implies:
\begin{equation}
\delta = \Upsilon\ \dot{u}\left(\tilde{p} \right)^{\frac{3}{4}}\,  \left(\frac{d \tilde{p}}{dx} \right)^{\frac{1}{2}}  + o\left(L^{-\frac{1}{2}} \right)
\end{equation}
Finally substituting this value in the free energy cost we recover its proportionality with $P\left(\tilde{p} \right)^{\frac{3}{4}}$ and the scaling found in the previous sections.

If on the other hand $A^{(4)}\left(Q \right) \neq 0$ we find the same scaling of the free energy cost with the number of spatial dimensions, but its dependence on $\tilde{p}$ does not reduce to a power of the overlap probability function.
 
\section{Conclusions}
In this work we extended the original FPV calculation in various directions, and at the same time checked the robustness of the original results. 
In the original works the scaling of the free energy density cost was obtained restricting the maximization procedure on a finite dimensional space of the whole solution space. The solution for different boundary conditions was assumed to be a small perturbation of the solution with free boundary conditions, the perturbations being second order polynomials in order to smooth discontinuities generated by the different boundary conditions.
Here we extended the solution space over which the maximization is performed to include higher order polynomials and continuity of higher order derivatives of the solution.
The space of solution thus explored is much larger than the original one, but still the solution originally found is robust: the scaling of the free energy cost as a function of the overlap difference and the number of spatial dimensions is independent of the smoothing we adopt for the discontinuities. 

We performed the same calculation in two extensions of the original model, to study the dependence of the free energy density cost on the overlap $q$ with different type of interactions. We found very different scenarios between an Hamiltonian with polynomial interactions and the one of the Edwards-Anderson model, characterized by interactions of the form $\texttt{Tr}\left(Q^n \right)$. 
In the case of interactions of the form $\sum_{ab}Q_{ab}^n$ we found a similar behaviour to the one of the hierarchical models of \cite{Franz2009} and of the reduced model originally analyzed in the FPV calculation: the free energy density cost depends on the overlap $q$ only through the function $P(q)$, and in particular it is proportional to $P(q)^{3/4}$. In the Edwards-Anderson model truncated to the fourth order, on the other hand, we found a much more complicated dependence on the overlap, not easily expressed in terms of physically meaningful functions.

To try and explain the different behaviours we evaluated analytically the free energy density cost for the generic Hamiltonian with interactions of the form $\sum_{ab}Q_{ab}^n$ and $\texttt{Tr}\left(Q^n \right)$. The main difference between the two classes of models is the ``locality'' of the interactions of order higher than three in replica space: while in the case of polynomial interactions all terms of order four and higher depend only on one value of the replica index, terms like $\texttt{Tr}\left(Q^4 \right)$ connect different replica indices. This has the effect of making the function $u\left(Q\right)$ explicitly dependent on $q_{max}$ and introducing more convoluted factors in the free energy cost dependence on the overlap.
The scaling of the free energy cost as a function of spatial dimensions is however not altered by these complications, which confirm the robustness of the FPV original calculation regarding the value of the lower critical dimension $D_{lc}= 2.5$.

\section{Acknowledgement}
This work was supported by a grant from the Simons Foundation (No. 454941, S. F.; No. 454949, G. P.), and from the European Research Council (ERC) under the European Union’s Horizon 2020 research and innovation programme (grant agreement No [694925]).

%This work was supported by grants from the Simons Foundation (No. 454941, S. F.; No. 454949, G. P.). This project has received funding from the European Research Council (ERC) under the European Union’s Horizon 2020 research and innovation programme (grant agreement No [694925]).

\begin{appendices}
\section{Eigenvalues}
One of the most useful tools used in the paper is the diagonalization of ultrametric matrices, sometimes called - in the continuum limit - Replica Fourier Transform \cite{crisanti2015replica}.
It is easy to prove that any $n$-dimensional hierarchical matrix of the form (here $n=8$):
\begin{equation}
Q = \begin{pmatrix}
q_d & q_2 & q_1 & q_1 & q_0 & q_0 & q_0 & q_0 \\
q_2 & q_d & q_1 & q_1 & q_0 & q_0 & q_0 & q_0 \\
q_1 & q_1 & q_d & q_2 & q_0 & q_0 & q_0 & q_0 \\
q_1 & q_1 & q_2 & q_d & q_0 & q_0 & q_0 & q_0 \\
q_0 & q_0 & q_0 & q_0 & q_d & q_2 & q_1 & q_1 \\
q_0 & q_0 & q_0 & q_0 & q_2 & q_d & q_1 & q_1 \\
q_0 & q_0 & q_0 & q_0 & q_1 & q_1 & q_d & q_2 \\
q_0 & q_0 & q_0 & q_0 & q_1 & q_1 & q_2 & q_d 
\end{pmatrix}
\end{equation}
can be written as $Q = U \Lambda_Q U^{-1}$ with $U$ a unitary matrix and
\begin{equation}
\Lambda_Q = q_d \, \mathbb{1} - \begin{pmatrix}
-\left( 4 q_0 + 2 q_1 + q_2\right) & 0 & 0 & 0 & 0 & 0 & 0 & 0 \\
0 & 4 q_0 - \left( 2 q_1 + q_2 \right) & 0 & 0 & 0 & 0 & 0 & 0 \\
0 & 0 & 2 q_1 - q_2 & 0 & 0 & 0 & 0 & 0 \\
0 & 0 & 0 & 2 q_1 - q_2 & 0 & 0 & 0 & 0 \\
0 & 0 & 0 & 0 & q_2 & 0 & 0 & 0 \\
0 & 0 & 0 & 0 & 0 & q_2 & 0 & 0 \\
0 & 0 & 0 & 0 & 0 & 0 & q_2 & 0 \\
0 & 0 & 0 & 0 & 0 & 0 & 0 & q_2 
\end{pmatrix}
\end{equation}
This can be obviously generalized for any matrix dimension $n$, and when taking the continuum limit $n\to 0$ \cite{mezard1987spin} - denoting as usual the matrix elements $Q(u)$ - the matrix eigenvalues can be written as:
\begin{equation}
\Lambda_{Q}(u) = q_d - \left(u \, Q(u)  + \int_u^1 dy\, Q\left(y \right) \right)
\end{equation}
In the paper we are interested mainly in two ultrametric matrices: $Q_{ab}$ and $P_{ab}$. In the following we list a number of algebraic properties of these matrices in the continuum limit, in their diagonal basis. 
In addition we are interested in the matrix $Q_{\alpha \beta}$ defined at the beginning of the paper. Being this matrix constructed in a hierarchical way from $Q_{ab}$ and $P_{ab}$ it is easy to write the eigenvalues of the former in terms of $\Lambda_Q$ and $\Lambda_P$. In particular we have:
\begin{equation}
Q_{\alpha \beta} = \begin{pmatrix}
Q_{ab} & P_{ab} \\
P_{ab} & Q_{ab}
\end{pmatrix}
\end{equation}
and being $Q$ and $P$ commuting matrices we obtain:
\begin{equation}
\Lambda_{Q_{\alpha \beta}} = \begin{pmatrix}
\left(\Lambda_{Q} +  \Lambda_{P}\right)_{ab} & 0 \\
0 & \left(\Lambda_{Q} - \Lambda_{P}\right)_{ab}
\end{pmatrix}
\end{equation}
While the matrix $P_{ab}$ has diagonal element $\tilde{p}$ the matrix $Q_{ab}$ has null diagonal by construction, so summing the eigenvalues of the two matrices we obtain (for the extrema of integration we use the same notation as in the paper):
\begin{equation}
\Lambda_{Q+P}(u) = \tilde{p} - \left[u (q+p)_u + \int_u^1 dy (q+p)_y  \right]
\end{equation}

\begin{equation}
\Lambda_{Q+P}(u) = \begin{cases}   - \left[2 u\, q_u + 2 \int_u^{\frac{u_p}{2}} dy\, q_y +  \int_{u_p}^{1} dy\, q_y  \right] \qquad & 0\leq u \leq \frac{u_p}{2}  \\  - \left[u_p\, \tilde{p} + \int_{u_p}^{1} dy\, q_y  \right] \qquad & \frac{u_p}{2} <  u  \leq u_p \\   - \left[u\, q_u + \int_u^1 dy\, q_y\right] \qquad & u_p <  u \leq 1   \end{cases}
\end{equation}
For their difference instead we have:
\begin{equation}
\Lambda_{Q-P}(u) = -\tilde{p} - \left[u (q-p)_u + \int_u^1 dy (q-p)_y  \right]
\end{equation}

\begin{equation}
\Lambda_{Q-P}(u) = \begin{cases}   - \left[u_p \tilde{p} + \int_{u_p}^1 dy\, q_y \right] \qquad & 0\leq u \leq u_p  \\  - \left[u\, q_u + \int_u^1 dy\, q_y  \right] \qquad & u_p <  u \leq 1   \end{cases}
\end{equation}
The same derivations performed for the matrices $Q_{\alpha \beta}$, $Q_{ab}$ and $P_{ab}$ can be carried over for their perturbations $\delta Q_{\alpha \beta}$, $\delta Q_{ab}$ and $\delta P_{ab}$:
\begin{equation}
\Lambda_{\delta Q + \delta P}(u) =  - \left[u (\delta q + \delta p)_u + \int_u^1 dy (\delta q + \delta p)_y  \right]
\end{equation}

\begin{equation}
\Lambda_{\delta Q + \delta P}(u) = \begin{cases}  0 \qquad & 0 <  u  \leq u_1 \\  - \left[2 u\, \delta q_u + 2 \int_u^{u_2} dy\, \delta q_y + \int_{u_3}^{u_4} dy\, \delta q_y  \right] \qquad & u_1 <  u \leq u_2  \\ - \int_{u_3}^{u_4} dy\, \delta q_y \qquad & u_2 <  u \leq u_3  \\
- \left[u\, \delta q_u  + \int_{u}^{u_4} dy\, \delta q_y  \right] & u_3 <  u \leq u_4 \\ 0 \qquad  & u_4 <  u \leq 1  \end{cases}
\end{equation}

\begin{equation}
\Lambda_{\delta Q - \delta P}(u) =  - \left[u (\delta q - \delta p)_u + \int_u^1 dy (\delta q - \delta p)_y  \right]
\end{equation}

\begin{equation}
\Lambda_{\delta Q - \delta P}(u) = \begin{cases} - \int_{u_3}^{u_4} dy\, \delta q_y  \qquad & 0\leq u \leq u_3    \\
- \left[u\, \delta q_u + \int_u^{u_4} dy\, \delta q_y  \right] & u_3 <  u \leq u_4 \\ 0 \qquad  & u_4 <  u \leq 1  \end{cases}
\end{equation}
Often we will also use the notation $\Lambda_{Q_u,\pm}$, $\Lambda_{\delta Q_u,\pm}$ to indicate $\Lambda_{Q \pm P(u)}$, $\Lambda_{\delta Q \pm \delta P(u)}$ in order to simplify the notation. 
In addition to the eigenvalues of the matrices $Q_{\alpha \beta}$ and $\delta Q_{\alpha \beta}$ for the evaluation of equation \eqref{trace} we need to compute integrals of the function $\delta q\left(u \right)$. It is however enough to compute these integrals at the leading order in $\delta$, so it is useful to express the function $\delta q_u$ up to the first order in $\delta$:

\begin{equation}
\delta q(u) = \frac{\delta}{4\ \dot{u}\left(\tilde{p} \right)}  \begin{cases}   0 \qquad & 0 \leq u < u_1 \\  -  \left(\frac{2u - u_p}{\delta} + 1 \right)^2  \qquad  & u_1 \leq u < \frac{u_p}{2}  \\ -  \left(\frac{2u - u_p}{\delta} - 1 \right)^2  \qquad & \frac{u_p}{2} \leq u < u_2   \\ 0  \qquad & u_2 \leq u < u_3 \\  \left(\frac{u - u_p}{\delta} + 1 \right)^2   \qquad & u_3 \leq u < u_p  \\  \left(\frac{u - u_p}{\delta} - 1 \right)^2   \qquad & u_p \leq u < u_5  \\0   \qquad & u_5 \leq u \leq 1 \end{cases}
\end{equation}
Finally we want to prove a set of useful identities for the integrals of perturbations in replica space, which will be widely used in the next appendix. The quantities of interest are:
\begin{equation}
I_{2,k} = \int_0^1 \frac{du}{u^2}\ (q_u - \tilde{p})^{2k-1} \left[ \Lambda_{\delta Q_u,+}^{2} + \Lambda_{\delta Q_u,-}^{2} \Theta(u - u_p)\right]
\end{equation}

Which, expanding the eigenvalues, become:
\begin{align}
I_{2,k} &= 4 \int_{\frac{u_p-\delta}{2}}^{\frac{u_p}{2}} du \ (q_u - \tilde{p})^{2k-1}  \left[ \delta q_u - \frac{1}{u} \int_{\frac{u_p - \delta}{2}}^{u} dy\, \delta q_y   \right]^{2} + \\
&\qquad \qquad \qquad \qquad \qquad + 2 \int_{u_p}^{u_p+\delta} du \ (q_u - \tilde{p})^{2k-1} \left[ \delta q_u  + \frac{1}{u} \int_{u}^{u_p + \delta} dy\, \delta q_y \right]^{2}  \nonumber
\end{align}
Without cancellations the leading order terms in the last expression would be proportional to $\delta^{2(k + 1)}$, but we have:
\begin{equation}
4 \int_{\frac{u_p-\delta}{2}}^{\frac{u_p}{2}} du \ (q_u - \tilde{p})^{2k-1}\ \delta q_u^{2} + 2 \int_{u_p}^{u_p+\delta} du \ (q_u - \tilde{p})^{2k-1}\ \delta q_u^{2}  = 0 
\end{equation}
so in the end we obtain:
\begin{equation}
\label{cancellation2}
I_{2,k} =  \frac{\Gamma_{k}\ \delta^{2k +3}}{u_p\ \dot{u}\left(\tilde{p}\right)^{2k +1 }} + o\left(\delta^{2(k+2)}\right)
\end{equation}
where $\Gamma_k$ are proportionality factors which will have to be evaluated performing the integrals.
\end{appendices}

\begin{appendices}
\section{Traces}
In this section we write in more details the derivation of the trace terms evaluated in section \ref{anres}.
For the first term we have:
\begin{eqnarray}
\label{firsttrace}
\texttt{Tr}\left(A^{(2)}(Q) \delta Q^2\right) &=&  -  \sum_{r=\{+,-\}} \int_0^1 \frac{du}{u} \frac{d}{du}\left(A^{(2)}(\Lambda_{Q_u,r})\Lambda_{\delta Q_u,r}^2  \right) \\
&=& \lim_{n\to 0}\frac{1}{n}A^{(2)}(\Lambda_{Q_n,-})\Lambda_{\delta Q_n,-}^2   + \sum_{r=\{+,-\}} \int_0^1 \frac{du}{u^2}\ A^{(2)}(\Lambda_{Q_u,r})\Lambda_{\delta Q_u,r}^2  \nonumber
\end{eqnarray}
%\begin{equation}
%\texttt{Tr}\left(A^{(2)}(Q)\delta Q^2\right) =  -\int_0^1 \frac{du}{u} \left[\dot{A}^{(2)}(\Lambda_{Q_u})\Lambda_{\delta Q_u}^2 + 2 A^{(2)}(\Lambda_{Q_u}) \Lambda_{\delta Q_u} \dot{\Lambda}_{\delta Q_u} \right]
%\end{equation}
%The last equation can be split in term of the single eigenvalues:
%\begin{equation}
%\label{singleeigeneq}
%\texttt{Tr}\left(A^{(2)}(Q)\delta Q^2\right) =  - \sum_{r=+,-} \int_0^1 \frac{du}{u} \left[\dot{A}^{(2)}(\Lambda_{Q_u,r})\Lambda_{\delta Q_u,r}^2 + 2 A^{(2)}(\Lambda_{Q_u,r}) \Lambda_{\delta Q_u,r} \dot{\Lambda}_{\delta Q_u,r} \right]
%\end{equation}
with $\Lambda_{Q_u,\pm}=\Lambda_{(q\pm p)_u}$ and $\Lambda_{\delta Q_u,\pm}=\Lambda_{(\delta q\pm \delta p)_u}$. Using the equations of motion we can write:
\begin{equation}
A^{(2)}\left( \Lambda_{(q + p)_u}\right) + B^{(2)}\left(q_u \right) = 0  %\qquad \forall\  q_u \neq \tilde{p}
\end{equation}
In addition we have: 
\begin{equation}
A^{(2)}\left( \Lambda_{(q - p)_u}\right) = \begin{cases} -  B^{(2)}\left(\tilde{p} \right) & q_u \leq \tilde{p} \\  - B^{(2)}\left(q_u \right)  \qquad & q_u > \tilde{p}   \end{cases}
\end{equation}
%It can be noticed that the integrand in \eqref{singleeigeneq} is different from zero only for $q_u \approx \tilde{p}$. The second term in the integral, involving derivatives of $\Lambda_{(\delta q\pm \delta p)_u}$, can be expressed as:
%\begin{eqnarray}
%\frac{1}{2} I_{\dot{\Lambda}_{\delta Q}} &=& \sum_{r=\{+,-\}} \int_0^1 \frac{du}{u}  B^{(2)}(q_u) \Lambda_{\delta Q_u,r} \dot{\Lambda}_{\delta Q_u,r}  \\
%&=& \sum_{r=\{+,-\}} \left[ \int_{\frac	{u_d - \delta}{2}}^{\frac	{u_d + \delta}{2}}  \frac{du}{u}   B^{(2)}(q_u) \Lambda_{\delta Q_u,r} \dot{\Lambda}_{\delta Q_u,r} + \int_{u_d - \delta}^{u_d + \delta}  \frac{du}{u}   B^{(2)}(q_u) \Lambda_{\delta Q_u,r} \dot{\Lambda}_{\delta Q_u,r} \right] \nonumber  \\
%&=&  - B^{(2)}(\tilde{p}) \texttt{Tr}\left(\delta Q^2\right) +  B^{(3)}(\tilde{p}) I_1 + B^{(4)}(\tilde{p}) I_2 + o\left( \delta^5 \right) \nonumber
%\end{eqnarray}
%We can prove another useful result:
%\begin{eqnarray}
%I_k &=& \sum_{r=\{+,-\}} \int_0^1 \frac{du}{u}  \left(q_u - \tilde{p}\right)^{2 k - 1} \Lambda_{\delta Q_u,r} \dot{\Lambda}_{\delta Q_u,r} = \\
%&=& 2 \Lambda_{\delta q}\left(u_2\right)\int_{u_p}^{u_p + \delta} du \left(q_u - \tilde{p}\right)^{2 k - 1} \dot{\delta q}(u) \nonumber
%\end{eqnarray}
Using the last identities equation \eqref{firsttrace} becomes:
\begin{eqnarray}
\texttt{Tr}\left(A^{(2)}(Q) \delta Q^2\right) &=& - B^{(2)}\left(\tilde{p} \right) \left[ \lim_{n\to 0}\frac{\Lambda_{\delta Q_n,-}^2}{n}   + \sum_{r=\{+,-\}} \int_0^1 \frac{du}{u^2}\ \Lambda_{\delta Q_u,r}^2\right] + \\
&-&  \sum_{j=1}^{\infty}\frac{B^{(j+2)}\left(\tilde{p} \right)}{j!} \int_0^1 \frac{du}{u^2}\ (q_u - \tilde{p})^j \left[ \Lambda_{\delta Q_u,+}^2 + \Lambda_{\delta Q_u,-}^2 \Theta(u - u_p)\right] = \nonumber  \\ 
&=&-  B^{(2)}\left(\tilde{p} \right)\texttt{Tr}\left(\delta Q^2\right) + I_{res,2} \nonumber 
\end{eqnarray}
with
\begin{equation}
I_{res,2} = - \sum_{j=1}^{\infty}\frac{B^{(j+2)}\left(\tilde{p} \right)}{j!} \int_0^1 \frac{du}{u^2}\ (q_u - \tilde{p})^j \left[ \Lambda_{\delta Q_u,+}^2 + \Lambda_{\delta Q_u,-}^2 \Theta(u - u_p)\right] 
\end{equation}
Barring accidental cancellations at order $\delta^5$ the only non-negligible terms in the last expression will be the ones proportional to $B^{(3)}\left(\tilde{p} \right)$ and $B^{(4)}\left(\tilde{p} \right)$. The contributions of both of them will be of order $\delta^5$, due to the cancellations proved in \eqref{cancellation2}.

We have for the dominant contributions:
\begin{align}
 I_{res,2} & =   \frac{4 B^{(3)}\left(\tilde{p} \right)}{u_p} \left[ \int_{\frac{u_p - \delta}{2}}^{\frac{u_p}{2}} du\ (q_u - \tilde{p})\ \delta q_u \left( \int_{\frac{u_p - \delta}{2}}^{u}dz\ \delta q_z\right)  -  \int_{u_p}^{u_p + \delta} du\ (q_u - \tilde{p})\ \delta q_u \left( \int_{u}^{u_p + \delta} dz\ \delta q_z \right)  \right] + \nonumber \\
& - B^{(4)}\left(\tilde{p} \right)  \left[ 2 \int_{\frac{u_p - \delta}{2}}^{\frac{u_p}{2}} du\ (q_u - \tilde{p})^2\ \delta q_u^2   +  \int_{u_p}^{u_p + \delta} du\ (q_u - \tilde{p})^2 \ \delta q_u^2   \right] + O\left( \delta^6 \right)
\end{align}
We will thus have (using \eqref{cancellation2} again) the result:
\begin{equation}
I_{res,2} = \left( \gamma \frac{B^{(3)}\left(\tilde{p} \right)}{u_p} + \zeta \frac{B^{(4)}\left(\tilde{p} \right)}{\dot{u}\left(\tilde{p} \right)}  \right) \frac{\delta^5}{\dot{u}\left(\tilde{p} \right)^{3}} + O\left(\delta^6 \right) 
\end{equation}
with $\gamma = \frac{27}{2520}$ and $\zeta = -\frac{1}{2520}$. This term is proportional to $\dot{u}\left(\tilde{p} \right)^{-3}$ if and only if $A^{(4)}\left(Q \right) = 0$, so we see a difference between the ``local" theories, in which the highest order trace term is $\texttt{Tr}\left(Q^3 \right)$, and theories like the Edwards-Anderson model.

The second trace term is given by:
\begin{eqnarray}
\label{secondtrace}
\texttt{Tr}\left(A^{(3)}(Q) \delta Q^3\right) &=&  -  \sum_{r=\{+,-\}} \int_0^1 \frac{du}{u} \frac{d}{du}\left(A^{(3)}(\Lambda_{Q_u,r})\Lambda_{\delta Q_u,r}^3  \right) \\
&=& \lim_{n\to 0}\frac{1}{n}A^{(3)}(\Lambda_{Q_n,-})\Lambda_{\delta Q_n,-}^3   + \sum_{r=\{+,-\}} \int_0^1 \frac{du}{u^2}\ A^{(3)}(\Lambda_{Q_u,r})\Lambda_{\delta Q_u,r}^3  \nonumber
\end{eqnarray}
For the third derivative of $A\left(\Lambda_Q \right)$ we have the identities:
\begin{equation}
A^{(3)}\left( \Lambda_{(q + p)_u}\right) = \begin{cases}   \frac{1}{2 u} B^{(3)}\left(q_u \right) & q_u < \tilde{p} \\  \frac{1}{u_p} B^{(3)}\left(\tilde{p} \right) & q_u = \tilde{p}   \\ \frac{1}{u} B^{(3)}\left(q_u \right)  \qquad & q_u > \tilde{p}   \end{cases}
\end{equation}

\begin{equation}
A^{(3)}\left( \Lambda_{(q - p)_u}\right) = \begin{cases}   \frac{1}{u_p} B^{(3)}\left(\tilde{p} \right) & q_u \leq \tilde{p}    \\ \frac{1}{u} B^{(3)}\left(q_u \right)  \qquad & q_u > \tilde{p}   \end{cases}
\end{equation}
As in the first trace term using these identities we obtain:
\begin{align}
& \texttt{Tr}\left(A^{(3)}(Q) \delta Q^3\right) = \frac{B^{(3)}\left(\tilde{p} \right)}{u_p} \texttt{Tr}\left(\delta Q^3\right) + I_{res,3} 
\end{align}
with
\begin{align}
I_{res,3} & =   \sum_{j=1}^{\infty}\frac{1}{j!}\left[ \left(\frac{d}{dq} \right)^j \frac{B^{(3)}\left(q \right)}{u\left( q \right)}\right]_{q=\tilde{p}} \times \\ & \times\int_0^1 \frac{du}{u^2}\ (q_u - \tilde{p})^j \left[\Lambda_{\delta Q_u,+}^3 \Theta( u_p - u) +  \left( \Lambda_{\delta Q_u,+}^3 + \Lambda_{\delta Q_u,-}^3\right) \Theta(u - u_p)\right]
\end{align}
where as usual $u\left( q \right)$ is the inverse function of $Q_f\left(u \right)$.
The only terms to analyze are $\texttt{Tr}\left(\delta Q^3\right)$ and the one in $I_{res,3}$ with $j=1$, the others giving subleading corrections. 
We find:
\begin{align}
\texttt{Tr}\left(\delta Q^3\right) & =  6 \left[  \int_{u_p - \delta}^{u_p + \delta} du\  \delta q_u^2 \left( \int_{u}^{u_p + \delta} dz\ \delta q_z \right) - 4 \int_{\frac{u_p - \delta}{2}}^{\frac{u_p + \delta}{2}} du\  \delta q_u^2 \left( \int_{\frac{u_p - \delta}{2}}^{u}dz\ \delta q_z\right)    \right] + \nonumber \\
& + 2  \left[ 2 \int_{\frac{u_p - \delta}{2}}^{\frac{u_p + \delta}{2}} du\ (2\ u - u_p)\ \delta q_u^3   +  \int_{u_p - \delta}^{u_p + \delta} du\ (u - u_p) \ \delta q_u^3   \right] + O\left(\delta^6 \right) = \nonumber \\
& = \eta \frac{\delta^5}{\dot{u}\left(\tilde{p}\right)^3} +  O\left(\delta^6 \right)
\end{align}
\begin{align}
I_{res,3} & =  \sigma\left(u_p \right) \int_0^1 \frac{du}{u^2}\ (q_u - \tilde{p}) \left[\Lambda_{\delta Q_u,+}^3\Theta( u_p - u) +  \left( \Lambda_{\delta Q_u,+}^3 + \Lambda_{\delta Q_u,-}^3\right) \Theta(u - u_p)\right] =  \nonumber \\
& = - 2\, u_p\,  \sigma\left(u_p \right) \left[ 4 \int_{\frac{u_p - \delta}{2}}^{\frac{u_p}{2}} du\ (q_u - \tilde{p}) \ \delta q_u^3   +  \int_{u_p}^{u_p + \delta} du\ (q_u - \tilde{p}) \ \delta q_u^3 \right] + O\left(\delta^6 \right) = \nonumber \\
& = \theta \frac{u_p\,  \sigma\left(u_p \right)}{\dot{u}\left(\tilde{p}\right)^4}\, \delta^5 +  O\left(\delta^6 \right)
\end{align}
with $\sigma(u) = \left( \frac{d}{dq}  \frac{B^{(3)}\left(q \right)}{u\left( q \right)}\right)_{q=Q_f(u)}$, $\eta = \frac{1}{20}$ and $\theta = -\frac{3}{1792}$. Finally for the trace we obtain:
\begin{equation}
\texttt{Tr}\left(A^{(3)}(Q) \delta Q^3\right) = \left( \eta \frac{B^{(3)}\left(\tilde{p} \right)}{u_p} + \theta\ \frac{u_p\, \sigma\left(u_p \right)}{\dot{u}\left( \tilde{p}\right)}  \right) \frac{\delta^5}{\dot{u}\left(\tilde{p} \right)^{3}} + O\left(\delta^6 \right) 
\end{equation}
As we noted previously the proportionality to $\dot{u}\left(\tilde{p} \right)^{-3}$ is attained if and only if $A^{(4)}\left(Q\right) = 0$.
The last significant term is the trace of fourth order, $ \texttt{Tr}\left(A^{(4)}(Q) \delta Q^4\right)$. To evaluate it we can use the identities:
\begin{equation}
A^{(4)}\left( \Lambda_{(q + p)_u}\right)  = - \begin{cases}   \frac{\sigma\left(u \right)}{2\, u} & q_u < \tilde{p} \\  \frac{\sigma\left(u_p \right)}{u_p} & q_u = \tilde{p}   \\ \frac{\sigma\left(u \right)}{u}  \qquad & q_u > \tilde{p}    \end{cases}
\end{equation}
\begin{equation}
A^{(4)}\left( \Lambda_{(q - p)_u}\right) = - \begin{cases}  \frac{\sigma\left(u_p \right)}{u_p} & q_u \leq \tilde{p}    \\ \frac{\sigma\left(u \right)}{u}  \qquad & q_u > \tilde{p}   \end{cases}
\end{equation}
The trace can then be expressed as:
\begin{align}
& \texttt{Tr}\left(A^{(4)}(Q) \delta Q^4\right) = - \frac{\sigma\left(u_p \right)}{u_p}  \left[ \lim_{n\to 0}\frac{\Lambda_{\delta Q_n,-}^4}{n}   + \sum_{r=\{+,-\}} \int_0^1 \frac{du}{u^2}\ \Lambda_{\delta Q_u,r}^4\right]   = \\
& = - 2\, u_p\, \sigma\left(u_p \right) \left[ 4 \int_{\frac{u_p - \delta}{2}}^{\frac{u_p + \delta}{2}} du\  \delta q_u^4   +   \int_{u_p - \delta}^{u_p + \delta} du\   \delta q_u^4   \right]  + O\left(\delta^6 \right) = \nonumber  \\ 
& =  \vartheta \frac{u_p\, \sigma\left(u_p \right)}{\dot{u}\left(\tilde{p} \right)^4 } \delta^5   + O\left(\delta^6 \right) \nonumber 
\end{align}
with $\vartheta = - \frac{1}{192}$.
\end{appendices} 

\bibliographystyle{plain}
\bibliography{StatMecField}

\end{document}